\documentclass[aps,jcp,reprint,nofootinbib,superscriptaddress]{revtex4-1}  
\usepackage{url}
\usepackage{breakurl}
\makeatletter
\g@addto@macro{\UrlBreaks}{\UrlOrds}
\g@addto@macro{\UrlBreaks}{\do\/\do\d}
\makeatother
\usepackage{graphicx}  
\usepackage{dcolumn}   
\usepackage{bm}        
\usepackage{amssymb,amsfonts,amsmath}   
\usepackage{tabu}
\usepackage{tabularx}
\usepackage{textgreek}
\usepackage{gensymb}
\usepackage{textcomp}
\usepackage{float}
\usepackage{placeins}
\usepackage[version=4]{mhchem}
\usepackage{hyperref}

\begin{document}
	
	\title{Kinetics of seeded protein aggregation: theory and application}

	\author{Alexander J. Dear}
	\email[]{alexander.dear@bc.biol.ethz.ch}
	\affiliation{Biochemistry and Structural Biology, Lund University, Sweden}
	\affiliation{Department of Biology, Institute of Biochemistry, ETH Zurich, Otto Stern Weg 3, 8093, Zurich, Switzerland}
	\affiliation{Bringing Materials to Life Initiative, ETH Zurich, Switzerland}
	
	\author{Georg Meisl}
	\affiliation{Centre for Misfolding Diseases, Department of Chemistry, University of Cambridge, Lensfield Road, Cambridge CB2 1EW, United Kingdom}
	
	\author{Jing Hu}
	\affiliation{Division of Physical Chemistry, Department of Chemistry, Lund University, Lund, Sweden}
	\affiliation{Biochemistry and Structural Biology, Lund University, Lund, Sweden}
	
	\author{Tuomas P. J. Knowles}
	\affiliation{Centre for Misfolding Diseases, Department of Chemistry, University of Cambridge, Lensfield Road, Cambridge CB2 1EW, United Kingdom}
	\affiliation{Cavendish Laboratory, University of Cambridge, J J Thomson Avenue, CB3 0HE, United Kingdom}
	
	\author{Sara Linse}
	\email[]{sara.linse@biochemistry.lu.se}
	\affiliation{Biochemistry and Structural Biology, Lund University, Sweden}
	
	\date{\today}

	\begin{abstract}
		``Seeding'' is the addition of preformed fibrils to a solution of monomeric protein to accelerate its aggregation into new fibrils. It is a versatile and widely-used tool for scientists studying protein aggregation kinetics, as it enables the isolation and separate study of discrete reaction steps contributing to protein aggregation, specifically elongation and secondary nucleation. However, the seeding levels required to achieve dominating effects on each of these steps separately have been established largely by trial-and-error, due in part to the lack of availability of integrated rate laws valid for moderate to high seeding levels and generally applicable to all common underlying reaction mechanisms. Here, we improve on a recently developed mathematical method based on Lie symmetries for solving differential equations, and with it derive such an integrated rate law. We subsequently develop simple expressions for the amounts of seed required to isolate each step. We rationalize the empirical observation that fibril seeds must often be broken up into small pieces to successfully isolate elongation. We also derive expressions for average fibril lengths at different times in the aggregation reaction, and explore different methods to break up fibrils. This paper will provide an invaluable reference for future experimental and theoretical studies in which seeding techniques are employed, and should enable more sophisticated analyses than have been performed to date.		
	\end{abstract}
	
	\pacs{87.14.em, 02.50.-r, 87.15.rp, 87.18.Tt}
	\maketitle
	
	
\section{Introduction}
Formation of amyloid fibrils from monomeric protein is a heavily-studied phenomenon that, when uncontrolled, drives a wide range of diseases in humans and animals~\cite{Walsh2020,Ankarcrona2016,Chiti2017,Westermark2005b}. When tightly regulated, it also plays functional roles in biology such as the formation of bacterial biofilms~\cite{Torkkeli2002,Linder2005,Lipke2023,Akbey2022}, and is beginning to be used in the construction of functional nanomaterials in industry~\cite{Kamada2021,Chowdhury2023,Han2023}.
	
A vital step toward preventing amyloid formation in disease, and toward the deliberate design of amyloid-based materials in industry, is to understand the molecular driving forces for amyloid formation as well as the underlying reaction mechanisms of the processes, and how these change from protein to protein and as a function of the reaction conditions. This can be achieved by coarse-grained kinetic modelling of experimental data, an approach which has met with much success in the past decade~\cite{Wang2011,Cohen2013,Meisl2014,Pilkington2017,Gaspar2017,Sang2018,Michaels2020,Dear2020PNAS,Camargo2021,Camargo2021b,Meisl2022func}. A key tool in studies of amyloid formation is the addition of varying amounts of preformed fibrils, ``seeds'', at the beginning of the reaction~\cite{Jarrett1992,Come1993,Jarrett1993,Jones2003,Orgel1996}. This terminology arose from the field of crystallography, in which it is common to add preformed ``seed'' crystals to promote new crystallization in a supersaturated solution~\cite{Cacciuto2004}. Its longstanding adoption in the field of amyloid fibrils is appropriate given that these fibrils share key characteristics in common with crystals. In particular, they have the periodic translational symmetry that is the defining feature of a crystal. Unlike traditional crystals, fibrils are highly anisotropic, consisting only of one or a few long filaments arranged in parallel and often twisted around one another; their periodic translational symmetry is therefore restricted primarily to the long axis. 

\begin{figure}
	\includegraphics[width=0.48\textwidth]{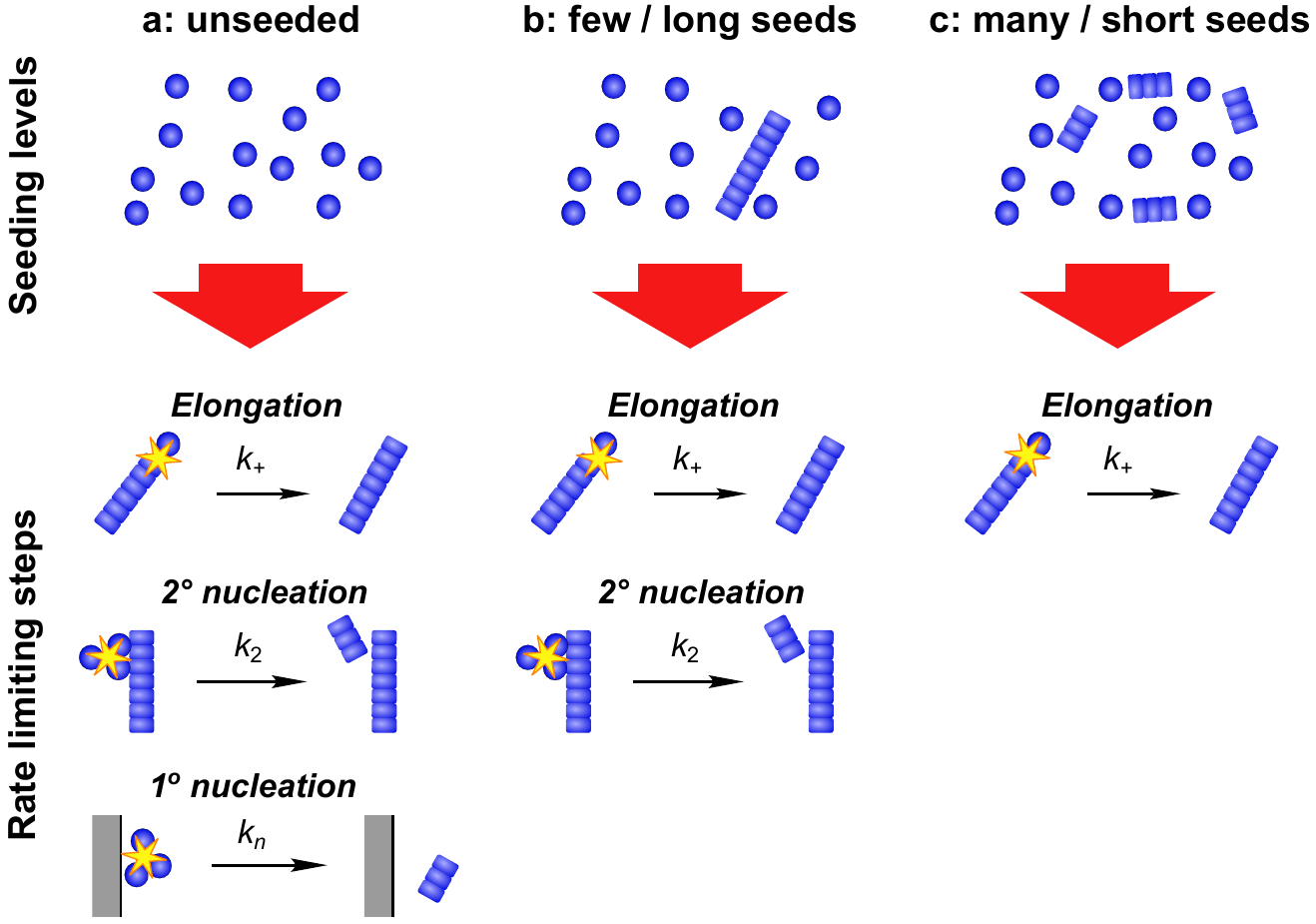}
	\caption{The key reaction steps in protein aggregation mechanisms and how they are affected by seed. \textbf{a}: Primary nucleation is the formation of fibrils from monomers in the absence of seed. \textbf{b}: With low initial seed concentrations, primary nucleation is not a significant source of new fibrils at any time. However, secondary nucleation remains responsible for the majority of new fibril formation over the entire reaction. \textbf{c}: At high seed levels, only elongation of existing fibrils is important for the overall kinetics.}
	\label{fig:intro}
\end{figure}

Data collected from seeded and unseeded reactions performed in parallel facilitate kinetic modelling of the amyloid formation process~\cite{Padrick2002,Meisl2016}. In particular, this is an effective way to study the different reaction steps in isolation (Fig.~\ref{fig:intro}), and how they are affected by changes in conditions such as the addition of a chaperone, or a change in temperature~\cite{Cohen2015,Cohen2018,Linse2020}. However, most available kinetic models~\cite{Knowles2009,Cohen2011a,Cohen2011b,Meisl2014,Dear2016,Michaels2016H,Dear2020JCP} have not been very precise when the initial seed concentration is moderate to large. Partly as a result, the amounts of seed required to isolate the different reaction steps have not been clearly established, and nor have ideal seed preparation techniques. As a result, seeding as a tool in kinetic studies has yet to live up to its full potential. 
	
The kinetics of the aggregation of monomeric protein (concentration $m(t)$) into fibrils \textit{in vitro}, where mass is conserved, can typically be accurately described by closed-form rate equations for the fibril number and mass concentrations $P(t)$ and $M(t)$~\cite{Michaels2014mf}. Since fibrils are long, these equations must include a (rapid) fibril elongation reaction step (Fig.~\ref{fig:intro}\textbf{a}-\textbf{c}). Primary nucleation of new fibrils from monomer is also obligate in the absence of seed (Fig.~\ref{fig:intro}\textbf{a}), and occurs in most \textit{in vitro} experiments even when seeded. Most systems studied \textit{in vitro} also feature an autocatalytic amplification reaction step (Fig.~\ref{fig:intro}\textbf{a}-\textbf{b}), most commonly secondary nucleation of new fibrils on the surfaces of existing fibrils~\cite{Ferrone1980,Ferrone1985b,Cohen2013}. Other possibilities include fibril breakage or fragmentation~\cite{Collins2004,Knowles2009}, or (unusually) fibril branching~\cite{Amann2001,Michaels2016H}.
	
Recently, a new mathematical method has been developed for finding accurate approximate solutions to an unusual class of differential equations~\cite{Dear2023B}. The rate equations describing a wide range of biofilament self-assembly reactions all fall into this class, opening up new possibilities for their kinetic modelling. The method involves the identification and exploitation of a type of symmetry property known as an asymptotic Lie symmetry possessed by the solutions to these equations.
	
Here, we generalize this method and use it to develop a highly accurate universal solution to the kinetics of protein aggregation with any amount of initial seed, valid for a range of commonly found reaction mechanisms. Using this solution, we derive a range of results relating to the effect of different seed concentrations on the kinetics. These results allow the calculation of the seeding levels required to isolate each reaction step, and how these levels depend on the overall mechanism of aggregation.

Sec.~\ref{sec:theory} presents the key theoretical results, including the general rate laws for seeded protein aggregation, their solutions, and analytical expressions for average fibril lengths during and after aggregation reactions. Readers interested primarily in experimental design should focus their attention mainly on Sec.~\ref{sec:experimental}, which explores the practical consequences of these results. Sec.~\ref{sec:mathmethods} contains the detailed mathematical methodology underpinning the study. Readers less interested in mathematical analysis of differential equations are advised to skip this section and simply use it as a reference for terminology, symbol definitions and equations where necessary. The Appendices contain various derivations of key results. They also contain further generalizations of the rate laws and their analytical solutions to incorporate other phenomena such as inhibitors.

\section{Theoretical results}\label{sec:theory}

\subsection{Rate equations for seeded and saturable protein aggregation kinetics}
	
We use standard terminology of $m(t)$ and $P(t)$ for monomer and fibril concentrations respectively. We also use $M(t)$ for fibril mass concentration: the concentration of the monomeric subunits that have been incorporated into fibrils ($M(t)$ is the key quantity reported on by Thioflavin T fluorescence experiments). At their simplest, the rates of elongation, primary nucleation and secondary processes are $\alpha_e=2k_+m$, $\alpha_1=k_nm(t)^{n_c}$ and $\alpha_2=k_2m(t)^{n_2}$, where $n_c$ and $n_2>1$. Secondary nucleation, as a catalytic step, obeys Michaelis-Menten-style kinetics and can saturate (saturation concentration $K_S$)~\cite{Meisl2014}. It has been shown that primary nucleation almost invariably occurs on interfaces such as the air-water interface or lipid membranes; thus, it too is often catalytic and can saturate at monomer concentration $K_P$~\cite{Sear2014,Espinosa2019,Pruppacher2010,Galvagnion2015,Dear2020JCP,Pham2016,Swasthi2017,Grigolato2017,Grigolato2021}. Finally, elongation can also saturate (at monomer concentration $K_E$) due to the finite residency time of the refolding monomer at the growing fibril end~\cite{Buell2010a}. In practice this usually occurs at higher concentrations than are typically used in experiments; nonetheless, it has been observed. These steps are therefore all well-described by Michaelis-Menten-type rate laws; this has been extensively validated against experimental data in the literature (see in particular \cite{Meisl2014,Buell2019,Dear2020JCP} for a detailed discussion of this and for further lists of references).
	
Putting this all together, the rate equations governing most protein aggregation reactions \textit{in vitro} are~\cite{Dear2020JCP}:
\begin{subequations}\label{momeqs}
	\begin{equation}\label{momeqP}
	\frac{dP}{dt}=\frac{k_n m(t)^{n_c}}{1+\left(m(t)/K_P\right)^{n_c}}+\frac{k_2 m(t)^{n_2}}{1+\left(m(t)/K_S\right)^{n_2}}M(t)
	\end{equation}
	\begin{equation}\label{momeqM}
	\frac{dM}{dt}=\frac{2k_+ m(t)}{1+m(t)/K_E}P(t)
	\end{equation}
	\begin{equation}
	m_\text{tot}=m(t)+M(t),
	\end{equation}
\end{subequations}
and the seed number and mass concentrations are $P(0)$ and $M(0)$. As usual, the negligible contributions of nucleation processes to the rate of increase of fibril mass concentration have been neglected~\cite{Michaels2014mf}, as have trivial contributions from processes such as fibril annealing (the inverse of fragmentation)~\cite{Michaels2014ann}. In practice, $P(t)$ and $M(t)$ include only fibrillar species, which corresponds to the species detected using the technique employed to collect experimental kinetic data (typically Thioflavin T fluorescence). Species such as solution-phase oligomers, which form as intermediates during nucleation of the detected fibrils, are not treated explicitly in these equations, which treat nucleation instead as a coarse-grained single-step rate law. It has been shown~\cite{Dear2020PNAS,Michaels2020,Michaels2022} that this coarse-graining does not lead to errors when only the fibril mass concentration is being tracked experimentally. The nucleation rate constants then contain contributions from both oligomer formation and oligomer conversion into elongation-competent fibrils. The model is agnostic as to how many filaments are in the fibrils, since this will not be distinguishable from kinetic curves of $M(t)$. The elongation rate constant $k_+$ can be viewed as an average of the rate constants for incorporation of new monomers into the growing fibril end for different individual filaments.

These rate equations with $K_S\to\infty$ can also describe fibril breakage or fragmentation ($n_2=0$)~\cite{Collins2004,Knowles2009}, or (unusually) fibril branching ($n_2=1$)~\cite{Amann2001,Michaels2016H}. (If primary nucleation saturation is present but is not modelled explicitly in the rate equations, it gives rise to $n_c$ values less than 1; this is the origin of the many instances of $n_c<1$ reported in the literature~\cite{Dear2024Nanoscale}.) Alternatively, in Appendix~\ref{app:competing} we consider rate laws that feature both fragmentation and secondary nucleation at the same time. In Appendix~\ref{app:inhib} we also derive more general rate laws that account for the effects of inhibitors binding monomers, oligomers, secondary nucleation sites, primary nucleation sites and fibril ends.

\subsection{Global analysis of kinetics}

The elongation rate is:
\begin{equation}\label{elrl}
\alpha_e(m)P=\frac{2k_+m}{1+m/K_E}P.
\end{equation}
The nondimensionalized monomer and fibril concentrations are $\mu(t)=m(t)/m_\text{tot}$ and 
\begin{equation}
\Pi(\tau)=\frac{2k_+P(\tau)}{\kappa(1+m_{\text{tot}}/K_E)},\label{Pnondim}
\end{equation}
where $\kappa$ will be defined later. As explained in Sec.~\ref{sec:asylims}, protein aggregation kinetics are largely governed by two asymptotic regimes, within which the kinetics simplify. Aggregation reactions start in a $(\mu\to 1,\ \Pi\to 0)$ asymptotic regime, defined by the validity of the perturbation solution Eq.~\eqref{genseries1}, and end in a $(\mu\to 0,\ \Pi\to\Pi_\infty)$ asymptotic regime, defined by validity of the linearized kinetics of Eq.~\eqref{mu3soln}. The latter regime is entered only when $\alpha_e\to 2k_+m_\text{tot}\mu$. When elongation is saturated, therefore, this regime is vanishingly small regardless of seeding levels. $\Pi_\infty$ is the final value of $\Pi$ at the end of an aggregation reaction, and is calculated in Appendix~\ref{sec:exactPi} for Eqs.~\eqref{momeqs}, giving Eq.~\eqref{Piinfty}.

\begin{figure}
	\includegraphics[width=0.48\textwidth]{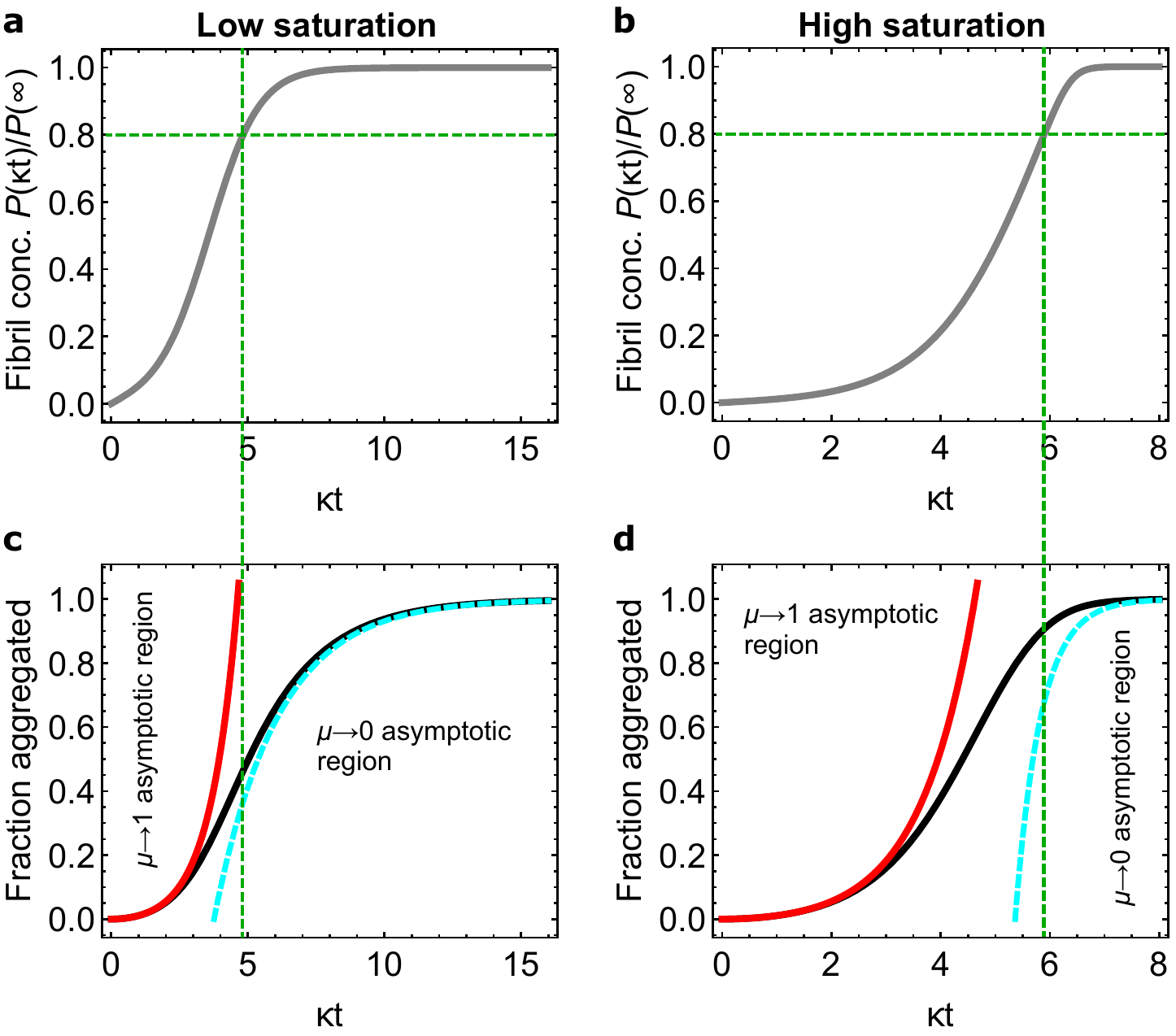}
	\caption{In the absence of elongation saturation, the kinetics of protein aggregation can be partitioned into two asymptotic regimes: $\mu\to 1$ and $\mu\to 0$. Parameters: $n_2=3,\ n_c=2,\ \varepsilon=0.01$. \textbf{a}-\textbf{b}: numerical trajectory for $\Pi$ can be successfully normalized to a maximum of 1 by dividing by $\Pi_\infty$ as calculated in Eq.~\eqref{Piinfty}, validating our analytical solution for $\Pi_\infty$. \textbf{a},\textbf{c}: Unsaturated kinetic curves ($K_S=10m_\text{tot}$) for $\Pi$ (numerical solution) and $M/m_\text{tot}$ (black: numerical solution). The point at which $\Pi=0.8\Pi_\infty$ is a good indicator of when we transition from the $\mu\to 1$ asymptotic regime for the kinetics to the $\mu\to 0$ regime. This is demonstrated by the $\mu\to 1$ asymptotic solution (Eq.~\eqref{genseries1}, red) losing accuracy and the  $\mu\to 0$ asymptotic solution (Eq.~\eqref{mu3soln}, cyan, dashed) becoming an accurate approximation to the overall kinetics. \textbf{b},\textbf{d}: Saturated ($K_S=0.04m_\text{tot}$) kinetic curves for $\Pi$ and $M/m_\text{tot}$. As expected, the transition between asymptotic regimes comes later, at lower $\mu$, due to the lower $\mu_c$ at which $\Pi$ stops growing significantly.}
	\label{fig:asyregions}
\end{figure}
	
When elongation is not saturated, the values of $\mu$ at which the $\mu\to 0$ regime is accessed, $\mu_c$, can be explored by solving $\Pi(\mu_c)=0.8\Pi_\infty$ (Fig.~\ref{fig:asyregions}\textbf{a}-\textbf{b}). For larger values of $n_2$, in the absence of saturation and small $\varepsilon$, we expect $\mu_c$ to increase, as secondary nucleation reduces more rapidly with decreasing $\mu$. Indeed, in the absence of seed $\mu_c(n_2=2)=0.43$ and $\mu_c(n_2=3)=0.56$ (Fig.~\ref{fig:asyregions}\textbf{c}). For saturating secondary nucleation we conversely expect much smaller $\mu_c$ values, as secondary nucleation remains important at much lower $\mu$ values. We also expect for small enough $K_S/m_\text{tot}$ that the dependence on $n_2$ is replaced by dependence on $K_S/m_\text{tot}$, as occurs with the expression for $\Pi_\infty$. This is borne out by explicit calculations: regardless of $n_2$, in the absence of seed $\mu_c(K_S=0.1m_\text{tot})=0.21$, and $\mu_c(K_S=0.01m_\text{tot})=0.045$. 

In summary, high seeding levels both increase the relative importance of the $\mu\to 0$ asymptotic regime and impact the kinetics in this regime. As explained in Sec.~\ref{sec:asylims}, this is both because the regime is entered into sooner, with $\mu$ starting from a smaller value, and because the seed concentrations affect the value of $\Pi_\infty$ directly. However, high saturation levels of secondary nucleation and of elongation can drastically reduce the importance of the $\mu\to 0$ asymptotic regime. Its importance is eliminated altogether when fragmentation occurs instead of secondary nucleation.

\subsection{General analytical solution to the kinetics of seeded, saturable protein aggregation}
The rates of primary and secondary nucleation are:
\begin{subequations}\label{prisecrl}
	\begin{align}
	\alpha_1(t,m)&=\frac{k_nm^{n_c}}{1+(m/K_P)^{n_c}}\\
	\alpha_2(m)M&=\frac{k_2m^{n_2}}{1+(m/K_S)^{n_2}}M.
	\end{align}
\end{subequations}
The dimensionless parameters $\kappa$ and $\varepsilon$ become:
\begin{subequations}
	\begin{align}
	\kappa&=\sqrt{\frac{2k_+k_2m_\text{tot}^{n_2+1}}{(1+m_{\text{tot}}/K_E)(1+(m_{\text{tot}}/K_S)^{n_2})}}\\
	\varepsilon&=\frac{k_nm_\text{tot}^{n_c}}{2k_2m_\text{tot}^{n_2+1}}\frac{1+(m_{\text{tot}}/K_S)^{n_2}}{1+(m_{\text{tot}}/K_P)^{n_c}}.\label{vareps}
	\end{align}
\end{subequations}
Their early-time equivalents are:
\begin{subequations}
	\begin{align}
	\kappa_0&=\sqrt{\frac{2k_+k_2m(0)^{n_2+1}}{(1+m(0)/K_E)(1+(m(0)/K_S)^{n_2})}}\\
	\varepsilon_0&=\frac{k_nm(0)^{n_c}}{2k_2m(0)^{n_2+1}}\frac{1+(m(0)/K_S)^{n_2}}{1+(m(0)/K_P)^{n_c}}.\label{vareps0}
	\end{align}
\end{subequations}
In Sec.~\ref{sec:mathmethods} a general solution, Eq.~\eqref{gensoln}, is presented for protein aggregation kinetics obeying rate laws of the form:
\begin{subequations}
	\begin{align}
	\frac{dP}{dt}&=\alpha_1(t,m)+\alpha_2(m)M(t)\\
	\frac{dM}{dt}&=\alpha_e(m) P(t),\quad m_\text{tot}=M(t)+m(t),
	\end{align}
\end{subequations}
where $\alpha_1(t,m)$, $\alpha_2(m)M$ and $\alpha_e(m)P$ are rates for primary nucleation (with possible explicit time-dependence), secondary processes and elongation respectively, with only mild restrictions placed on their forms. These restrictions are all met by the rates describing seeded and saturated protein aggregation (given by Eqs.~\eqref{elrl} and \eqref{prisecrl}). The solution for the rate equations, Eqs.~\eqref{momeqs}, describing seeded and saturated protein aggregation can therefore be calculated by substituting these specific forms for $\alpha_1$, $\alpha_2$ and $\alpha_e$ into Eq.~\eqref{gensoln} and simplifying, ultimately giving:
\begin{widetext}
	\begin{subequations}\label{gensolnsat}
	\begin{align}
	m(t)&=m(0)\left(1+\frac{p_0}{2c}\frac{\kappa_0}{\kappa}(e^{\kappa t}-e^{-\kappa t})+\frac{\kappa_0^2}{\kappa^2}\left(\frac{\varepsilon_0}{c}+\frac{\delta_0}{2c}+\frac{p_0^2}{2c^2}\left(1+\frac{c\,m(0)/K_E}{1+m(0)/K_E}\right)\right)(e^{\kappa t}+e^{-\kappa t}-2)\right)^{-c},\\
	\delta_0&=\frac{M(0)}{m(0)}, \qquad p_0=\frac{2k_+P(0)}{\kappa_0(1+m(0)/K_E)}.\label{p0}
	\end{align}
\end{subequations}
%
Sec.~\ref{sec:mathmethods} offers two choices for the formula for $c$, depending on whether the parameters $p$, $\varepsilon$ and $\delta$ affect the overall kinetics most strongly via their influence on the $\mu\to 0$ or the $\mu\to 1$ asymptotic regime. If secondary nucleation or elongation is saturated, or if fragmentation occurs instead of secondary nucleation, the $\mu\to 0$ asymptotic regime is negligible in extent and these parameters affect the overall kinetics much more strongly in the $\mu\to 1$ asymptotic regime. $c$ is then best chosen according to the $\mu\to 1$ asymptotic symmetry transformation formula Eq.~\eqref{cmu1}, which for these rate equations is:
\begin{equation}\label{muto1c}
c=\frac{3}{2n_2'+1},\quad n_2'=\frac{n_2}{1+(m_{\text{tot}}/K_S)^{n_2}}-\frac{2m_{\text{tot}}/K_E}{1+m_{\text{tot}}/K_E}.
\end{equation}
Otherwise, $c$ is best chosen according to the $\mu\to 0$ asymptotic symmetry transformation formula Eq.~\eqref{cmu0}, which for these rate equations depends on $\Pi_\infty$ via Eq.~\eqref{Piinfty}. In this limit, Eq.~\eqref{Piinfty} simplifies greatly, and Eq.~\eqref{cmu0} becomes:
\begin{equation}\label{Piinftyprisat}
	c=\left(p^2+\frac{4\varepsilon}{n_c}\left(1+\left(\frac{K_P}{m_\text{tot}}\right)^{n_c}\right)\ln\!\left[1+\left(\frac{m(0)}{K_P}\right)^{n_c}\right]+2\left(\frac{\mu_0^{n_2}}{n_2}-\frac{\mu_0^{1+n_2}}{1+n_2}\right)\right)^{1/2},\quad\mu_0=\frac{m(0)}{m_\text{tot}},\quad p=\frac{2k_+P(0)}{\kappa}.
\end{equation}
(See Sec.~\ref{sec:nontechoverview} for an overview of asymptotic symmetry transformations and Appendix~\ref{app:asymptoticmethod} for their use to derive Eqs.~\eqref{gensoln}, \eqref{cmu1} and \eqref{cmu0}.) In the case that both fragmentation and secondary nucleation occur at comparable rates, which has occasionally been observed~\cite{Cohen2013}, this solution can be straightforwardly generalized to account for this (see Appendix~\ref{app:competing}). In Appendix~\ref{app:inhib} we generalize it even further to account for the effects of inhibitors binding monomers, oligomers, secondary nucleation sites, primary nucleation sites and fibril ends.

\end{widetext}

\begin{figure}
		\centering\includegraphics[width=0.48\textwidth]{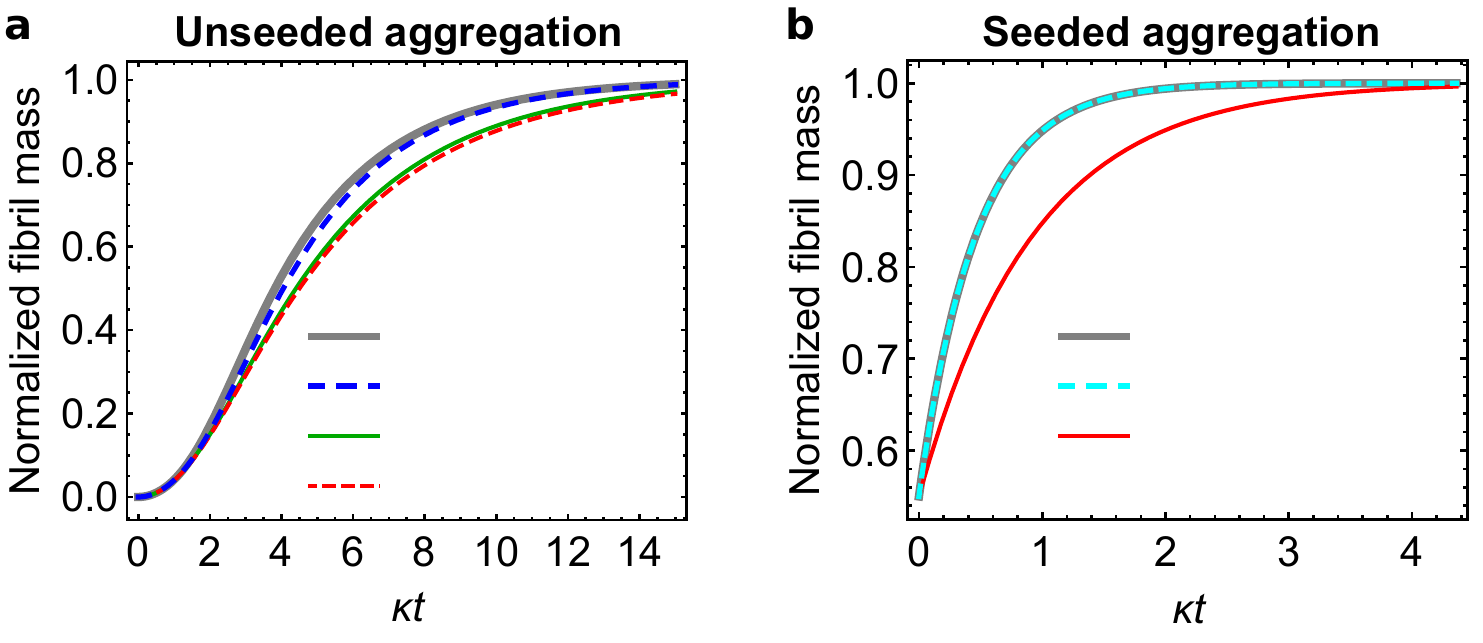}
		\put(-57,32.5){\fontfamily{phv}\selectfont\tiny Solution from~\cite{Cohen2011b}}
		\put(-57,41){\fontfamily{phv}\selectfont\tiny Eq.~\eqref{gensolnsat}}
		\put(-57,49){\fontfamily{phv}\selectfont \tiny Numerical solution}
		\put(-181,24){\fontfamily{phv}\selectfont\tiny Solution from~\cite{Michaels2016H}}
		\put(-181,32.5){\fontfamily{phv}\selectfont\tiny Solution from~\cite{Michaels2019b}}
		\put(-181,41){\fontfamily{phv}\selectfont\tiny Eq.~\eqref{gensolnsat}}
		\put(-181,49){\fontfamily{phv}\selectfont \tiny Numerical solution}
		\caption{Comparison of Eq.~\eqref{gensolnsat} to earlier solutions for the kinetics of linear protein self-assembly. \textbf{a}: Eq.~\eqref{gensolnsat} offers a modest improvement in accuracy over earlier solutions when describing unseeded kinetics with no fibrils present at $t=0$. Parameters: $n_2=5$, $n_c=3$, $\varepsilon=0.04$ and no saturation. \textbf{b}: Earlier solutions fail to describe adequately the kinetics when moderate concentrations of fibrils are present at $t=0$. By contrast, Eq.~\eqref{gensolnsat} provides almost exact results. Parameters: $n_2=5$, $n_c=4$, $\varepsilon=0.1$, $\Pi(0)=2.16$, $\mu(0)=0.45$ and no saturation.}
		\label{fig6}
	\end{figure}

This solution performs better for unseeded kinetics than previous solutions (Fig.~\ref{fig6}\textbf{a}), giving confidence in its theoretical underpinnings. More importantly, where earlier solutions break down at high seed, Eq.~\eqref{gensoln} remains accurate (Fig.~\ref{fig6}\textbf{b}). This combination of accuracy and generality makes it attractive for kinetic data fitting purposes.
	
\subsection{Saturation greatly affects average fibril lengths}\label{results:avelength}
	
	\begin{figure*}
		\includegraphics[width=0.96\textwidth]{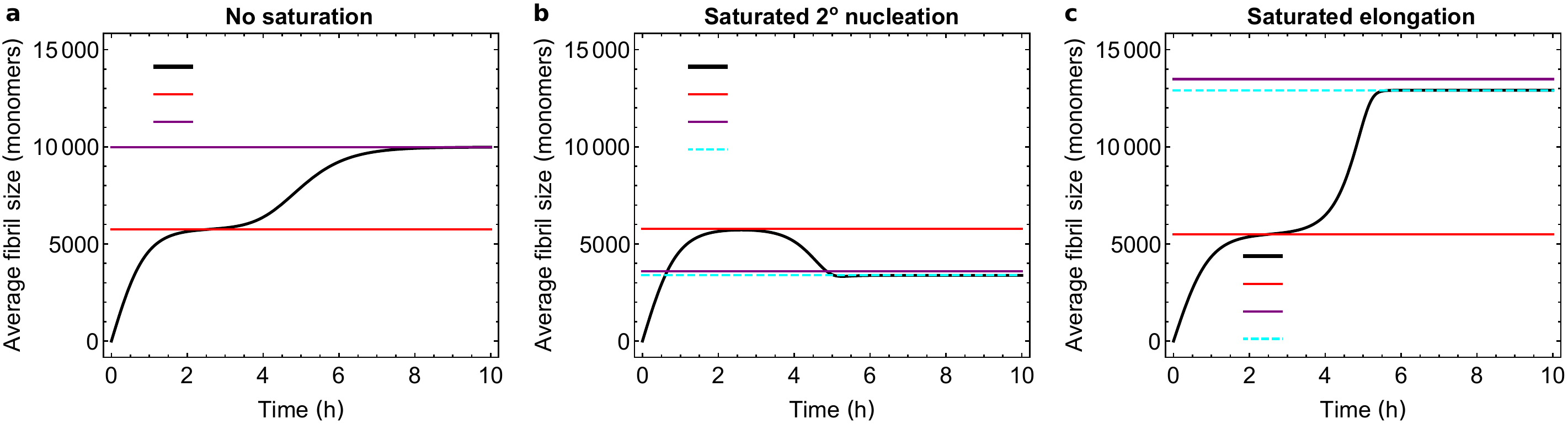}
		\put(-84,50){\fontfamily{phv}\selectfont \tiny Numerical solution to $L(t)$}
		\put(-84,42){\fontfamily{phv}\selectfont \tiny Steady-state length $L_\text{\tiny ss}$, Eq.~\eqref{eq:lss}}
		\put(-84,33.5){\fontfamily{phv}\selectfont \tiny Approximate $L_\infty$, Eq.~\eqref{Linf_elsat}}
		\put(-84,25){\fontfamily{phv}\selectfont \tiny Exact $L_\infty$, Eqs.~\eqref{Linfty}-\eqref{Piinfty}}
		\put(-258,110){\fontfamily{phv}\selectfont \tiny Numerical solution to $L(t)$}
		\put(-258,101.5){\fontfamily{phv}\selectfont \tiny Steady-state length $L_\text{\tiny ss}$, Eq.~\eqref{eq:lss}}
		\put(-258,93){\fontfamily{phv}\selectfont \tiny Approximate $L_\infty$, Eq.~\eqref{Linf_secsat}}
		\put(-258,84.5){\fontfamily{phv}\selectfont \tiny Exact $L_\infty$, Eqs.~\eqref{Linfty}-\eqref{Piinfty}}
		\put(-425,110){\fontfamily{phv}\selectfont \tiny Numerical solution to $L(t)$}
		\put(-425,101.5){\fontfamily{phv}\selectfont \tiny Steady-state length $L_\text{\tiny ss}$, Eq.~\eqref{eq:lss}}
		\put(-425,93){\fontfamily{phv}\selectfont \tiny Exact $L_\infty$, Eq.~\eqref{Linf_nosat}}
		\caption{The evolution of average fibril lengths (number of monomeric subunits) over time is well captured by our analytical solutions and has a complex dependence on the reaction mechanism. Initially new fibrils are formed mainly through primary nucleation, and the average length always increases with time. Once the fibril mass becomes high enough that secondary processes take over as the main fibril formation mechanism, the average fibril length achieves steady-state, given by Eq.~\eqref{eq:lss}. \textbf{a}: When the mechanism is unsaturated ($K_X=100m_\text{tot}$ for all steps) and features secondary nucleation, the fibrils start to increase in length from the steady-state value towards final value Eq.~\eqref{Linf_nosat} once significant depletion of monomer begins to occur. This is because fibril formation reduces more rapidly with monomer concentration than elongation. \textbf{b}: Conversely, saturation in secondary nucleation ($K_S/m_\text{tot}=0.1$) causes fibril lengths to reduce after the steady-state length has been attained, since now elongation is more sensitive to monomer depletion. For the same $v_\text{max}$, the final average fibril length (Eq.~\eqref{Linf_secsat}) is thus lower. \textbf{c}: Saturation in elongation ($K_E/m_\text{tot}=0.1$) causes a more pronounced growth in fibril length upon monomer depletion, reaching a higher final value (Eq.~\eqref{Linf_elsat}), as the relative sensitivity of secondary nucleation to falling monomer levels becomes more pronounced. Parameters: A\textbeta40 reaction orders and rate constants~\cite{Meisl2014}, modified to ensure the same $v_\text{max}=k_xK_X^{n_x}$ in each panel as in~\cite{Meisl2014}. Longer times spent at $L_\text{ss}$ are in principle achieved in systems with smaller $\varepsilon$.}
		\label{fig:avelength}
	\end{figure*}
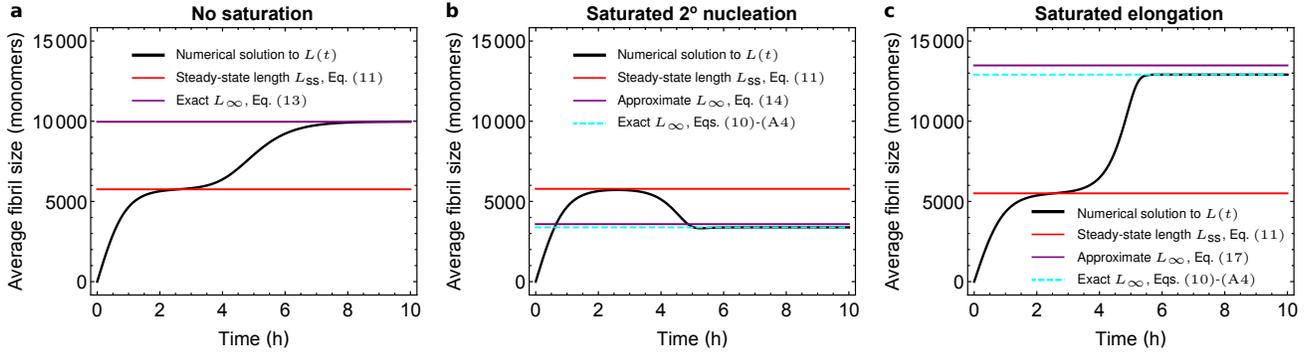
	
We can use Eq.~\eqref{Pnondim} to express the average length of fibrils at the end of an aggregation reaction (or equivalently in the $\mu\to 0$ asymptotic limit), $L_\infty=m_\text{tot}/P_\infty$, in terms of $\Pi_\infty$:
\begin{equation}\label{Linfty}
	L_\infty=\frac{2k_+m_\text{tot}}{\kappa\Pi_\infty(1+m_\text{tot}/K_E)}.
\end{equation}
Assuming monomer solubility is low, $\Pi_\infty$ can be computed from $\Pi(\mu)$ (Eq.~\eqref{Pimu}) by setting $\mu\to 0$, yielding Eq.~\eqref{Piinfty}. 
	
This is to be compared to the steady-state average fibril length in the $\mu\to 1$ asymptotic regime, $L_{ss}$. This is equivalent to when monomer has not yet appreciably depleted, but when enough time has elapsed for the kinetics in this regime (Eqs.~\eqref{genseries1}) to be dominated by the time-exponential term with the positive exponent. $L_{ss}$ is then given by dividing $1-$ Eq.~\eqref{mu11} by Eq.~\eqref{Pi11}, taking the late-time limit, and redimensionalizing, yielding:
\begin{equation}
	L_{ss}=\frac{2k_+m_\text{tot}}{\kappa(1+m_\text{tot}/K_E)}.
	\label{eq:lss}
\end{equation}
So, $L_\infty=L_{ss}/\Pi_\infty$.
	
In the absence of all saturation, Eq.~\eqref{Piinfty} reduces to (see Appendix~\ref{app:amylofit}):
\begin{equation}\label{Piinftynosat}
	\Pi_\infty=\left(p^2+\frac{4\varepsilon\mu_0^{n_c}}{n_c}+2\left(\frac{\mu_0^{n_2}}{n_2}-\frac{\mu_0^{1+n_2}}{1+n_2}\right)\right)^{1/2}.
\end{equation}
In the absence of initial seed, Eq.~\eqref{Linfty} then reduces to:
\begin{equation}
	L_\infty=\frac{2k_+m_\text{tot}}{\kappa\sqrt{2/(n_2(n_2+1))+4\varepsilon/n_c}}.\label{Linf_nosat}
\end{equation}
Since for secondary nucleation $n_2>1$, fibrils get longer on average as the reaction approaches completion (Fig.~\ref{fig:avelength}\textbf{a}).
	
In the limit that only secondary nucleation is saturated, the functional form of $L_{ss}$ is unchanged although $\kappa$ changes. However, $L_\infty$ becomes (see Appendix~\ref{app:avelength}):
\begin{equation}
	L_\infty=\frac{2k_+m_\text{tot}}{\kappa\sqrt{p^2+2\ln\!\left[m(0)/K_S\right]-2\mu_0}}.\label{Linf_secsat}
\end{equation}
We see that $L_\infty$ becomes much smaller than $L_{ss}$ when secondary nucleation is saturated (Fig.~\ref{fig:avelength}\textbf{b}). Note $L_\infty$ has lost its dependence on $n_2$ in this limit.
	
Because in reality $\Pi_\infty=\Pi(\mu=m_s/m_\text{tot})$, where $m_s$ is the solubility limit, the fibril number concentration does not in practice grow indefinitely with $m_\text{tot}/K_S$. Once $K_S<m_s$ (but still assuming $m_s\ll m_\text{tot}$), $\Pi_\infty$ is instead given by:
\begin{equation}
	\Pi_\infty\simeq\left(p^2+2\ln\!\left[\frac{m(0)}{m_s}\right]-2\mu_0\right)^{1/2}.
\end{equation}
At this point the correction for saturation to the average seed length no longer depends on $K_S$.
	
Finally we look at the case of saturation in elongation only ($m_\text{tot}/K_E\gg 1$). Now $\Pi_\infty$ reduces to:
\begin{equation}
	\Pi_\infty=\left(p^2+\frac{4\varepsilon\mu_0^{1+n_c}}{1+n_c}+2\left(\frac{\mu_0^{1+n_2}}{1+n_2}-\frac{\mu_0^{2+n_2}}{2+n_2}\right)\right)^{1/2},
\end{equation}
and we see we have gained an increase of 1 in $n_c$ and $n_2$ relative to the unsaturated case. This reflects the fact that the rates of primary and secondary nucleation relative to that of elongation reduce more rapidly with decreasing $\mu$ when elongation is saturated, by an extra power of $\mu$. Consequently, the fibrils get longer (Fig.~\ref{fig:avelength}\textbf{c}). Without seed, $L_\infty$ becomes:
\begin{equation}
	L_\infty=\frac{2k_+m_\text{tot}/(1+m_\text{tot}/K_E)}{\kappa\sqrt{2/((n_2+1)(n_2+2))+4\varepsilon/(n_c+1)}}.\label{Linf_elsat}
\end{equation}
	
\section{Implications for experimental design}\label{sec:experimental}
Seeded kinetic experiments can be used to help separate out the contribution of different reaction steps to the overall proliferation of amyloid fibrils. This is of particular value when attempting to isolate the effect of changing conditions on the different steps or when determining which steps are affected by an inhibitor or accelerator of aggregation, as it allows one to switch off the contribution from specific processes. The theory developed in the preceding sections can be used to determine with greater precision the seeding levels required to do this. In this section we outline how this can be done in practice.
	
\subsection{Seeding levels required to bypass solely primary nucleation}
The critical seed concentration at which primary nucleation produces as many new fibrils as secondary nucleation is given by the ratio of primary to secondary nucleation rates:
\begin{equation}
	M_\text{crit,1}=2m(0)\varepsilon_0=\frac{k_nm(0)^{n_c}}{k_2m(0)^{n_2}}\frac{1+(m(0)/K_S)^{n_2}}{1+(m(0)/K_P)^{n_c}}.
\end{equation}
As a quick rule of thumb, to eliminate the impact of primary nucleation on the aggregation kinetics we choose a seed concentration 10 times higher than the critical concentration $M(0)>10M_\text{crit,1}$, because this ensures that secondary nucleation produces an order of magnitude more new fibrils than primary nucleation. (A more rigorous justification of this rule follows below.) For A\textbeta42 and A\textbeta40 under typical \textit{in vitro} reaction conditions, $\varepsilon_0$ is very small and yields nanomolar concentrations for $M_\text{crit,1}$, so primary nucleation can be bypassed at very low seed concentrations already. Clearly, when a system does not possess appreciable secondary nucleation, primary nucleation cannot be bypassed in favour of secondary nucleation.
	
The elimination of the impact of primary nucleation on the aggregation kinetics has several uses. First, by comparison with analysis of unseeded kinetics, it allows investigation of the effect of changes in conditions on specifically primary nucleation~\cite{Linse2020}. Second, the amount of seed required to achieve this depends on the ratio of the rate of primary nucleation to that of secondary processes, and has thus long been used as a diagnostic for the presence of the latter~\cite{Meisl2016,Meisl2022func}. If only very small percentages of seed are required to greatly accelerate the kinetics, this can only be because $\varepsilon$ is very small, and the kinetics thus dominated by secondary nucleation, fragmentation, or branching. This effect has been demonstrated in great detail by Arosio et al. \cite{Arosio2015}. Third, since Eq.~\eqref{gensolnsat} implies that the reaction half-time depends logarithmically on seed concentration, fluctuating levels of seed impurities have far less of an impact on the kinetics when a concentration of seed $\gg M_\text{crit,1}$ has already been deliberately added (i.e.\ $\ln(10\varepsilon_0+s)$ varies much less strongly with $s$ than $\ln(\varepsilon_0+s)$). 
	
The above rule of thumb can be arrived at more rigorously by considering the form of Eq.~\eqref{gensolnsat}, which reveals that once $M(0)/m(0)+p_0\kappa/\kappa_0\gg 2\varepsilon_0$, the kinetics are independent of primary nucleation at all times. If the seeds were generated in an unseeded but otherwise identical reaction (i.e.\ same initial monomer concentration), then $P(0)=M(0)/L_\infty$, and Eq.~\eqref{Linfty} can be used to express $p_0$ as $p_0=\Pi_\infty(\text{unseeded}) M(0)/m(0)$. Since in unseeded reactions with small $\varepsilon$ (secondary-nucleation-dominated kinetics) $\Pi_\infty\kappa/\kappa_0$ is generally $O(1)$, this condition can be rewritten as:
\begin{equation}
    \frac{M(0)}{m(0)}\gg \varepsilon_0 = \frac{M_\text{crit,1}}{2m(0)}.
\end{equation}
    
\begin{figure*}
	\includegraphics[width=0.7\textwidth]{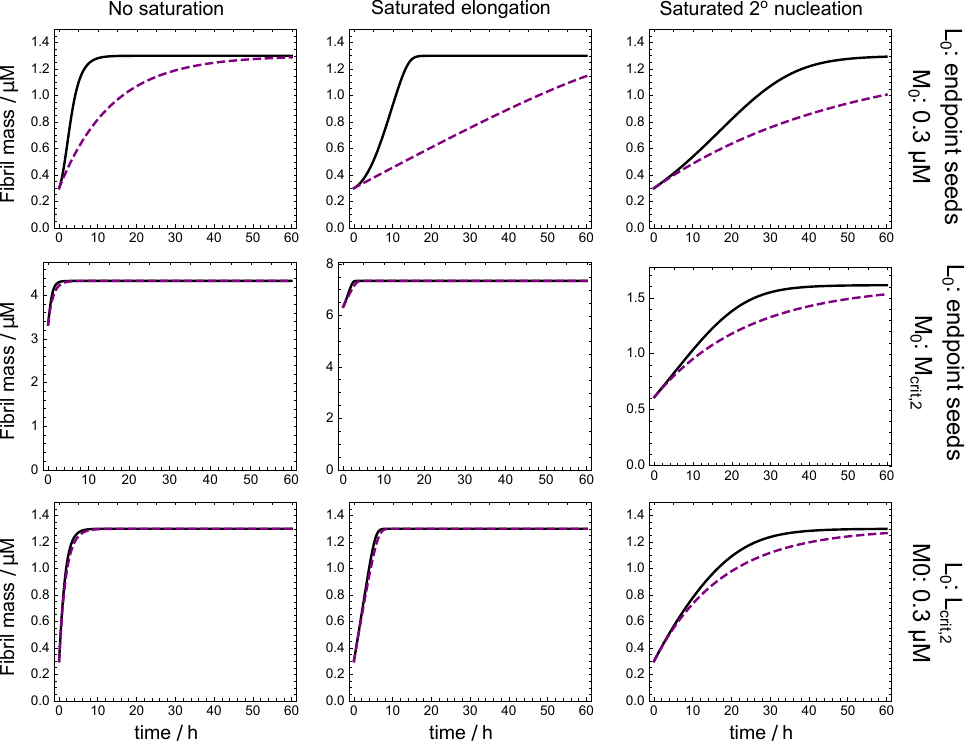}
	\caption{Comparing the the aggregation kinetics with secondary nucleation (black) with a hypothetical reaction without secondary nucleation (purple, dashed), for a number of different seed lengths and seed concentrations. Kinetic curves are simulated by numerical integration of Eqs.~\eqref{momeqs}. Top row: 30\% seed (0.3 $\mu$M), length of fibrils is that present at the end of a reaction starting from the same monomer concentration, without seed. Middle row: seed concentration is $M_{crit,2}$, length of fibrils is that present at the end of a reaction starting from the same monomer concentration, without seed. Bottom row: 30\% seed (0.3 $\mu$M), length of fibrils is $L_{crit,2}$. The left column assumes no saturation, the middle column saturation of elongation only, and the right column saturation of secondary nucleation only. Parameters used: $m_0 = 1$, $k_+ = 10^4$, $k_2 = 3.6\cdot10^{-5}$ (black) or $k_2=0$ (purple, dashed), $n_2 = 2$, $K_E = 0.1$ (relevant only in middle column) and $K_S = 0.1$ (relevant only in right column), with units in $\mu$M and h in all cases.
	}
	\label{fig:Elon_only}
\end{figure*}

\subsection{Seeding levels required to bypass both primary and secondary nucleation}
	
Adding enough seed allows us to investigate specifically the effect of changing reaction conditions on elongation~\cite{Meisl2016}, such as changing temperature~\cite{Cohen2018}, buffer ionic strength~\cite{Meisl2017b}, or adding potential inhibitory drugs~\cite{Shammas2011,Linse2020,Aprile2017b,Cox2018}. This follows because when $P(0)$ is large enough, $P(t)$ is at all times already close enough to $P_\infty$ that increases in $P(t)$ due to the nucleation of new fibrils do not significantly affect the overall kinetics for $m(t)$. These are thus well-described by the linearized kinetics in the late-time (or $\mu\to 0$) regime, shown in Sec.~\ref{sec:mathmethods} to be given by Eq.~\eqref{mueqred3}. By comparison to the effect of these changes on an unseeded reaction or a reaction with low seed, their impact on the other reaction steps such as secondary nucleation can subsequently be inferred.
	
As a quick rule of thumb, for the kinetics to be dominated by fibril elongation, seed concentrations should be well above the critical concentration (derived in Appendix~\ref{app:M0crit}):
\begin{equation}
    M_\text{crit,2}\approx m_0\left(\frac{L_0^2}{2L_{ss0}^2}\right)\left(1+\sqrt{1+
	\frac{4}{e}\frac{L_{ss0}^2}{L_0^2 }}\right),
    \label{equ:Mcrit2}
\end{equation}
where $L_0$ is the average seed length, $m_0$ the initial monomer concentration and $L_{ss0}=\sqrt{2k_+m_0/\alpha_2(m_0)}$ the steady-state length that would be achieved if the reaction proceeded without seed fibrils (becoming e.g.\ $L_{ss0}=\sqrt{2k_+m_0/(k_2m_0^{n_2})}$ when secondary nucleation is unsaturated). If one is using seeds directly from the plateau of a reaction under similar conditions, this usually means seed concentrations should be higher than monomer concentrations to push the system to a regime where secondary processes are irrelevant. Note Eq.~\eqref{equ:Mcrit2} was derived assuming no saturation in elongation.
	
One notable feature of Eq.~\ref{equ:Mcrit2} is the scaling of $M_{crit,2}$ with seed length: At low values of the initial length, $L_0$, compared to the steady state length, $L_{ss0}$, the scaling is linear, i.e. decreasing seed length is as effective as increasing seed concentration at shifting towards a more elongation dominated regime. However, at larger initial seed lengths, the scaling of $M_{crit,2}$ with $L_0$ becomes quadratic, so in this regime decreasing seed length is significantly more effective than increasing seed concentrations. This effect is illustrated in more detail in Fig.~\ref{fig:SI_Mcrit_performance_figure}.
	
A similar calculation can be performed for systems in which only primary nucleation, but not secondary nucleation is present, giving $M_{crit}=m_0\sqrt{\frac{n_c}{2}}\frac{L_0}{L_{\infty 0}}$ (see again Appendix~\ref{app:M0crit}). Therefore, the required level of seeding to bypass all nucleation is similar with or without secondary processes, but when there are no secondary processes its dependence on seed length is always linear.
    
One can alternatively determine the seed \textit{length} required to bypass all nucleation for a given seed \textit{concentration}, $L_{crit,2} = \sqrt{\frac{M_0^2L_{ss0}^2}{m_0(m_0+M_0e)}}$ (also derived in Appendix~\ref{app:M0crit}). Unsurprisingly, a lower seed concentration requires the average seed length to be reduced, to ensure there are still enough growth-competent fibril ends for elongation to dominate. Both strategies are used in Fig.~\ref{fig:Elon_only} to demonstrate how a system can be shifted to an elongation dominated regime by either increasing seed concentration or changing seed length. Both strategies have also been used in practice. Although using high concentrations of seed to study elongation has been the most common approach, especially for A\textbeta~\cite{Meisl2016,Yang2018,Cohen2018}, sonication has also been used deliberately to shorten seed fibrils to facilitate isolation of the elongation step in kinetic assays~\cite{Milto2013,Milto2014,Sakunthala2022,Adam2024}. 
	
\subsection{Practical approaches to better analyze fibril elongation}
	
The normal seeding approach to isolate the elongation reaction step has been to add enough seed to eliminate the lag phase, which for A\textbeta42 and others is generally held to be around 30\% seed/monomer concentration ($\delta\simeq 0.23$). However, the above calculations reveal that such seeding levels can be insufficient to achieve this (see Fig.~\ref{fig:Elon_only}). They also reveal that the seeding levels required to fully isolate the elongation reaction step, using seed fibrils taken from the endpoint of an unseeded reaction under the same conditions, are impractical. This is because very high seed levels lead to issues in accurately detecting the increase in fibril concentration via fluorescence, and because reactions will be very fast, potentially too fast to measure. Factors like temperature equilibration lead to a loss in accuracy of kinetic measurements at early times, thus causing problems when a significant part of the reaction takes place during these early times.
	
Unless the same seed stock is used and one is interested only in the relative changes of the elongation rate across different conditions, the initial seed numbers have to be determined to then calculate the absolute elongation rate. To do so requires first that the average seed fibril length $L_0$ be calculated, and second that this is used to compute $P(0)=M(0)/L_0$, because $k_+$ enters the highly-seeded kinetics always multiplied by $P(0)\simeq P_\infty$ (equivalent to the $\mu\to 0$ kinetics Eq.~\eqref{mueqred3}, as explained above). In practice measuring the average length of seed fibrils is often impossible, as they are often much larger than the field of view in cryo-TEM, and often clumped together into hard-to-interpret tangles. Even when neither of these issues feature, the fibril size distribution is still often quite broad, making these average measurements more challenging.
	
These issues all indicate that the more reliable approach for studying elongation is to reduce the length of the seed fibrils, provided fibril structure can be conserved. As revealed by Eq.~\eqref{equ:Mcrit2} and Fig.~\ref{fig:Elon_only}, by greatly increasing $P(0)$ it is possible to reach the elongation-dominated regime at a much lower $M(0)$. This slows greatly the secondary nucleation rate and drops greatly the baseline ThT signal, both of which lead to much more accurate kinetics. Moreover, a smaller $M(0)$ and shorter seeds reduces issues related to clumping or gelation, prevents fibrils being longer than the field of view in the microscopy images, and reduces the breadth of the size distribution,  in principle permitting more accurate length measurements. 
	
In principle, seed length can be reduced by varying the conditions under which seeds are made, for example changing the monomer concentration will change the average length of fibrils at the end of the reaction (Eq.~\eqref{Linfty}). However, it may be difficult to predict changes in fibril length accurately and changing conditions may also modify other aspects such as fibril morphology. Therefore, the best strategy is to break up the seed fibrils into much smaller pieces prior to starting the seeded aggregation reaction. This has the additional advantage of allowing more efficient usage of seeds from a rare or expensive source.
	
\begin{figure}
	\includegraphics[width=0.48\textwidth]{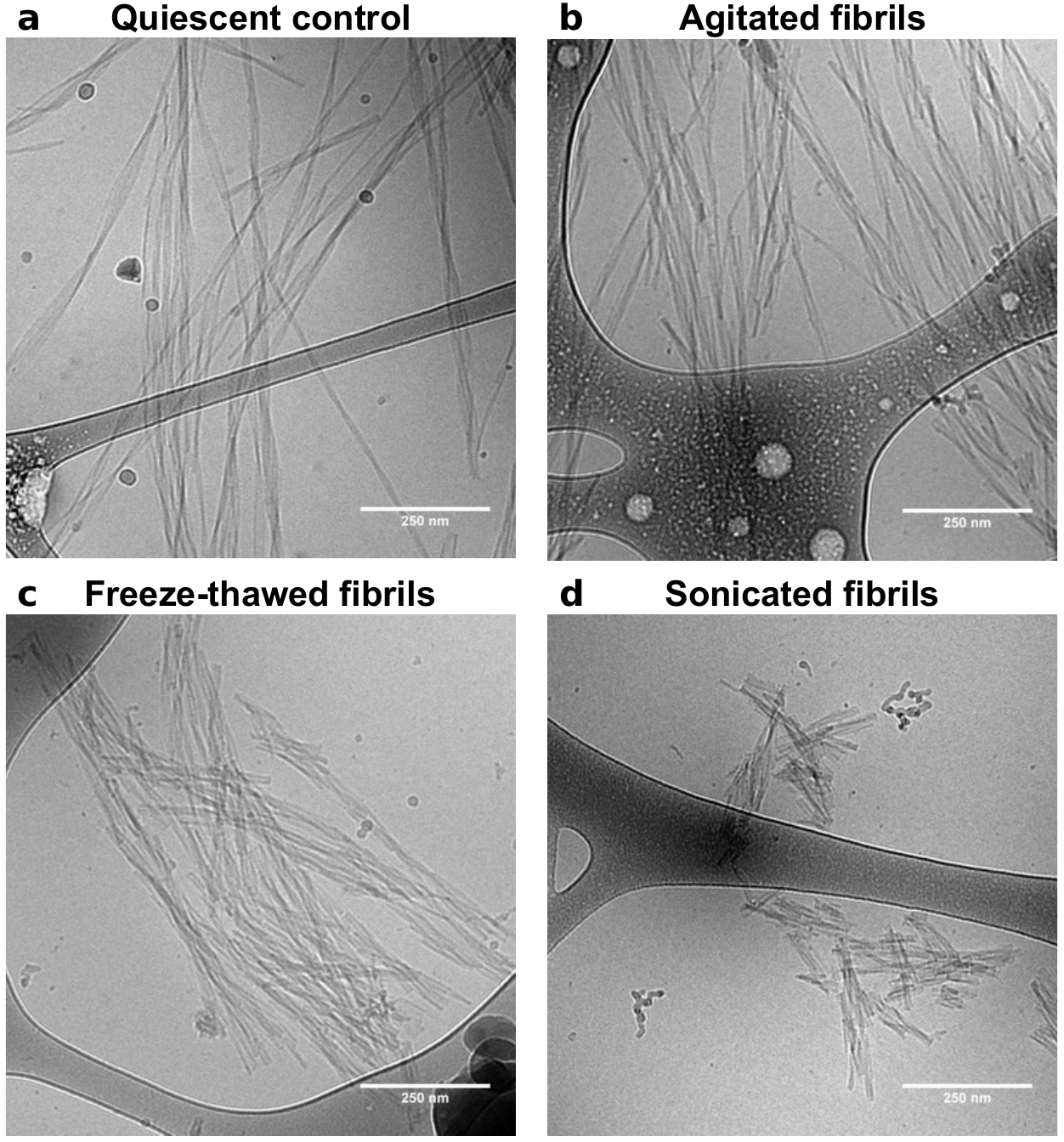}
	\caption{Comparison of different methods for breaking up A\textbeta40 fibril seeds using cryo-TEM microscopy. \textbf{a}: Untreated fibril seeds are hundreds of nm in length. \textbf{b}: Shaking the seeds at 2000 rpm for 1 hour at 37 \celsius\ has little visible effect on seed length or morphology. \textbf{c}: Fibril seeds appear shorter after 10 freeze/thaw cycles from -80 to +37 \celsius. \textbf{d}: Tip sonication appears much more effective at breaking A\textbeta40 fibrils, with the cryo-TEM images showing much shorter fibrils. Neither of these treatments appears to affect fibril morphology in terms of fibril width and node-to-node distance.}
	\label{fig:cryo}
\end{figure}
	
\begin{figure}
	\includegraphics[width=0.48\textwidth]{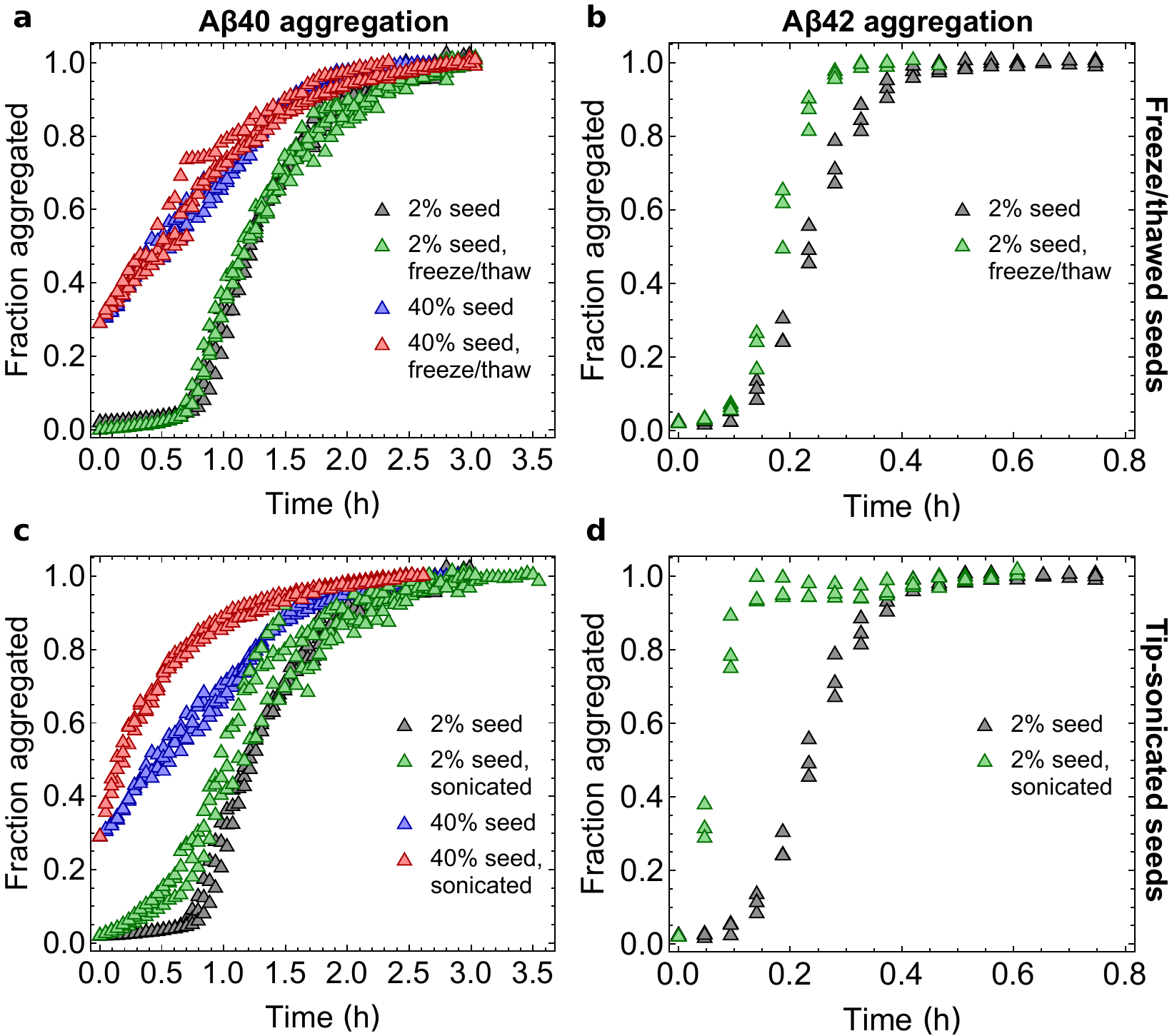}
	\caption{Comparison of different methods for breaking up fibril seeds, by their ability to accelerate the aggregation of A\textbeta40 (6 \textmu M) and A\textbeta42 (3 \textmu M). \textbf{a}: Freeze/thaw treatment (10 cycles from -80 to +37 \celsius) has no detectable effect on the seeding propensity of A\textbeta40 fibrils, implying A\textbeta40 fibrils cannot be appreciably fragmented by this treatment. \textbf{b}: Conversely, the same treatment induces a small but measurable acceleration in the seeding propensity of A\textbeta42 fibrils, implying they are more sensitive to this treatment. \textbf{c-d}: Tip sonication is much more effective at breaking fibrils of both types, as evidenced by the notable increase in their seeding propensity. Again, A\textbeta40 fibrils (\textbf{c}) are less sensitive to the treatment compared with A\textbeta42 fibrils (\textbf{d}), implying they are mechanically stronger. The seed \% is given relative to initial monomer concentration.}
	\label{fig:breaking}
\end{figure}
	
There are a few potential methods for achieving this breaking-up of seed fibrils into shorter segments. We tested these on A\textbeta40 and A\textbeta42 fibrils, which have historically not usually been deliberately broken up prior to seeding. We first tried mechanical agitation on A\textbeta40 fibrils (see Methods), an approach that is known to break A\textbeta42 fibrils~\cite{Cohen2013}. However, this made no difference to the length or appearance of A\textbeta40 fibril seeds under cryo-TEM (Fig.~\ref{fig:cryo}\textbf{a}-\textbf{b}), and also caused no increase in seeding propensity of A\textbeta40 fibrils, implying that this method does not successfully break up A\textbeta40 fibrils. Second, we tested the effect of repeated cycles of freezing and thawing on A\textbeta40 and A\textbeta42 fibrils (see Methods), which might be expected to break up fibrils due to the shear forces induced by ice crystal formation. However, although this appears to reduce somewhat the length of A\textbeta40 fibril seeds (Fig.~\ref{fig:cryo}\textbf{c}), it has no effect at all on their seeding propensity (Fig.~\ref{fig:breaking}\textbf{a}), and only modestly increases A\textbeta42 fibril seeding propensity (Fig.~\ref{fig:breaking}\textbf{b}). The apparent reduction in A\textbeta40 seed length clearly demonstrates the bias of cryo-TEM towards visualization of the shorter, more dispersed fraction of fibrils, with the great majority of A\textbeta40 fibrils being contained in large un-analyzable clumps or tangles with little change in their length.
	
Finally and most importantly, sonication has traditionally been used to break up fibrils into smaller fragments~\cite{Cannon2004,Buell2012b,Milto2013,Milto2014}. Different fibril types are known to require different amounts of sonication to achieve a given average length~\cite{Buell2012b}. We tested this explicitly for A\textbeta40 and A\textbeta42 fibrils, using the same sonication settings for both (see Methods). cryo-TEM indicated a more significant shortening of A\textbeta40 fibril seeds (Fig.~\ref{fig:cryo}\textbf{d}). In agreement with this, their seeding propensity was increased by a measurable although not large amount, with an aggregation reaction containing 2\% sonicated A\textbeta40 seeds still possessing a lag phase (Fig.~\ref{fig:breaking}\textbf{c}). Seeding at 29\% revealed sonication to increase the initial slope (and thus the rate of elongation) by 2-3x, consistent with a 2-3x reduction in average A\textbeta40 fibril length by sonication. By contrast, the seeding propensity of the A\textbeta42 fibrils was greatly increased by the same sonication treatment (Fig.~\ref{fig:breaking}\textbf{d}), with only 2\% of the sonicated seeds already sufficient to eliminate the lag phase.
	
Together, these findings demonstrate that it is harder to break A\textbeta40 fibrils than it is to break A\textbeta42 fibrils. This is despite the CAC for A\textbeta40 being higher, implying that A\textbeta40 fibrils are \textit{less} thermodynamically stable, not more. This can be understood by considering the differences in fibril structures. A\textbeta40 fibrils are known to be broader in cross-section, containing at least 4 filaments, whereas A\textbeta42 fibrils contain only two filaments under the \textit{in vitro} conditions used here~\cite{Schmidt2009}. To disaggregate a single monomer from the end of the fibrils requires breaking contacts to the same number of monomeric subunits in each case, with the interaction strength between subunits being weaker in the case of A\textbeta40 fibrils. However, fragmenting a fibril requires breaking at least double the number of bonding interactions in A\textbeta40 fibrils than in A\textbeta42 fibrils. That this is harder to do tells us that the increase in number of bonds that must be broken outweighs the individually weaker bonds and requires more energy overall to achieve.
	
\section{Conclusions}
	
We have made innovations to the method of asymptotic Lie symmetries and used these to derive a highly accurate and general integrated rate law for protein aggregation with arbitrary number and mass concentrations of seed, and arbitrary amounts of saturation in any of the reaction steps. Previous integrated rate laws were not capable of accurately describing the kinetics in the presence of moderate or high seed concentrations, regardless of whether any reaction steps are saturated. We have updated online protein aggregation kinetic model-fitting program AmyloFit with this more general solution, which supersedes several less accurate earlier integrated rate laws. We have also derived expressions for average fibril lengths produced when different reaction steps saturate. Together these fill some of the remaining gaps in our understanding of ``standard'' protein aggregation reactions.
	
Using these analytical results we quantified how much seed is required to be added to a reaction to effectively bypass the primary nucleation reaction step. Although this type of seeded reaction is often performed, the required seeding level has previously always been guessed. Here we instead provided mathematical proof for the diagnostic role small amounts of seed can provide for the presence of secondary processes. The exact seed concentration required will depend on all rate constants in a system:
\begin{equation}
\frac{M(0)}{m(0)}> 20\varepsilon_0.
\end{equation}
We can for example state that for X \textmu M total A\textbeta42 monomer, X nM seed would be required to bypass primary nucleation.
	
We also investigated the seed levels typically used to bypass all nucleation to study elongation in isolation, and found these to be often insufficient. We further showed that without modification of the fibrils, the seed concentrations required to reliably bypass all nucleation are often impractically high. Instead, we showed that sonication of seed fibrils is often the best approach to study elongation in isolation, although the amount of sonication required scales with the number of filaments contained in the amyloid fibrils. Although the exact seed concentration required will depend on the length of the seed fibrils and all rate constants in a system, we can for example state that with X \textmu M A\textbeta42 monomer, X nM of seed would be required to bypass all nucleation.
	
Taken together, we believe the kinetics of seeded protein aggregation, and the various types of seeding reactions frequently performed, have finally been put on a rigorous footing, and that this paper will provide a reference for scientists performing such experiments in the future.

\section{Mathematical methods}\label{sec:mathmethods}

Here, we write down general rate laws for a wide variety of protein aggregation reactions and perform an asymptotic analysis on them to reveal their characteristic behaviour. We also outline qualitatively the method used to generate global solutions to these equations.
	
\subsection{General rate equations and their nondimensionalization}\label{sec:rateeqs}
To maximize the applicability of our analysis, we consider as general a class of rate equations as possible, those considered in~\cite{Dear2023B}:
\begin{subequations}\label{MP1}
	\begin{align}
	\frac{dP}{dt}&=\alpha_1(t,m)+\alpha_2(m)M(t)\label{P1}\\
	\frac{dM}{dt}&=\alpha_e(m) P(t),\quad m_\text{tot}=M(t)+m(t).\label{M1}
	\end{align}
\end{subequations}
$\alpha_2M$ and $\alpha_eP$ are general expressions for the rates of secondary processes and of elongation, and $\alpha_1(t,m)$ is a general rate law for primary nucleation processes, depending on time $t$ both explicitly and implicitly via $m(t)$. Since elongation involves monomer addition, $\lim_{m\to m_c} \alpha_e=2k_+(m-m_c)$ must always hold, where $m_c\ll m_\text{tot}$ is the equilibrium concentration for elongation (and is usually approximated as zero). Neglecting reverse nucleation reactions, the rates $\alpha_1,\ \alpha_2$ are never negative. 

We furthermore require that the rates $\alpha_2$ and $\alpha_e$ depend on constant parameters $\bm{d}$ in such a way that $\bm{d}=0$ reduces them to $\alpha_2(m,\bm{d}=0)=k_2m^{n_2}$, and $\alpha_e(m,\bm{d}=0)=2k_+m$. This ensures the existence of a special solution when $\bm{d}=0$, which is necessary for treatment of these equations by Lie symmetry methods~\cite{Dear2023B}. This is quite a gentle restriction and encompasses most known rate laws including saturation of elongation and secondary nucleation~\cite{Dear2020JCP}, and inhibition~\cite{Michaels2020I}.
	
Eqs.~\eqref{MP1} were nondimensionalized and simplified in~\cite{Dear2023B}, yielding:
\begin{subequations}\label{geneqsnondim}
	\begin{align}
	\frac{d\Pi}{d\tau}&=2\varepsilon\frac{\alpha_1(t,m)}{\alpha_1(0,m_\text{tot})}+\frac{\alpha_2(m)}{\alpha_2(m_\text{tot})}(1-\mu(\tau))\label{geneqnondimP}\\
	\frac{d\mu}{d\tau}&=-\frac{\alpha_e(m)}{\alpha_e(m_\text{tot})} \Pi(\tau).\label{geneqnondimm}
	\end{align}
\end{subequations}
The dimensionless variables are:
\begin{subequations}
    \begin{align}
        \tau&=\kappa t\\
        \mu&=\frac{m}{m_\text{tot}}\\
        \Pi&=\frac{\alpha_e(m_\text{tot}) P}{m_\text{tot}\kappa}.
    \end{align}
\end{subequations}
The key dimensionless parameters are:
\begin{subequations}
    \begin{align}
    \kappa &=\sqrt{\alpha_e(m_\text{tot})\alpha_2(m_\text{tot})}\\
    \varepsilon&=\frac{\alpha_1(0,m_\text{tot})}{2m_\text{tot}\alpha_2(m_\text{tot})}.
    \end{align}
\end{subequations}
$\kappa$ is the fundamental timescale of protein aggregation via secondary processes, and $\varepsilon$ can be interpreted as the relative importance of primary nucleation over secondary processes. The initial conditions in nondimensional form are then:
\begin{subequations}\label{ICs}
	\begin{align}
	\mu(0)&=\mu_0=1-\delta,\quad\delta=\frac{M(0)}{m_\text{tot}}\\
	\Pi(0)&=p=\frac{\alpha_e(m_\text{tot})P(0)}{\kappa m_\text{tot}}.
	\end{align}
\end{subequations}
For $\varepsilon=\bm{d}=0$ and $p=p^*$, Eqs.~\eqref{geneqsnondim} were solved in~\cite{Dear2023B}, giving:
	\begin{subequations}\label{refsoln}
		\begin{align}
		\mu^*(\tau,c,\delta)&=\frac{1}{\left(1+e^\tau\left[(1-\delta)^{-1/c}-1\right]\right)^{c}},\\
		c&=\frac{3}{2n_2+1}\\
		p^*=&\sqrt{2\frac{1-(1-\delta)^{n_2}}{n_2}-2\frac{1-(1-\delta)^{n_2+1}}{n_2+1}}.
		\end{align}
	\end{subequations}
Note, $p^*=\delta+O(\delta^2)$.
	
\subsection{$\mu\to 1$ asymptotic limit of dynamics}\label{sec:muto1}

Eqs.~\eqref{geneqsnondim} cannot be solved exactly for $\mu$ and therefore $M$. However, they can be expanded perturbatively in $\delta,\ p$ and $\varepsilon$ in this regime by multiplying each of these parameters by a bookkeeping parameter $s$, to be later set to 1, that is used as the single perturbation parameter~\cite{Dear2023B}. The substitutions $\mu=\sum_{i=0}s^i\mu^{(i)}$ and $\Pi=\sum_{i=0}s^i\Pi^{(i)}$ are then made. To $O(s^0)$ the equations are easily verified to be solved by $\mu^{(0)}=1,\ \Pi^{(0)}=0$. To $O(s^1)$ the perturbation equations are:
\begin{subequations}\label{perteqsO1}
	\begin{align}
	\frac{d\Pi^{(1)}}{d\tau}&=2\varepsilon\frac{\alpha_1(t,m_\text{tot})}{\alpha_1(0,m_\text{tot})}-\mu^{(1)}(\tau)\\
	\frac{d\mu^{(1)}}{d\tau}&=- \Pi^{(1)}(\tau).
	\end{align}
\end{subequations}
Solving these and combining with the $O(s^0)$ solutions yields the following overall solutions for $\mu$ and $\Pi$ to $O(s)$~\cite{Dear2023B}:
\begin{subequations}\label{genseries1}
	\begin{align}
	\Pi(\tau)&=s\!\left[\varepsilon\dot{\mathcal{F}}(\tau)+\frac{\delta}{2}(e^\tau-e^{-\tau})+\frac{p}{2}(e^\tau+e^{-\tau})\right],\label{Pi11}\\
	\mu(\tau)&=1-\,s\!\left[\vphantom{\frac{p}{2}}\varepsilon\mathcal{F}(\tau)\right.\nonumber\\
	&\qquad\qquad\quad\left.+\frac{\delta}{2}(e^\tau+e^{-\tau})+\frac{p}{2}(e^\tau-e^{-\tau})\right],\label{mu11}
	\end{align}
\end{subequations}
where $\mathcal{F}(\tau)$ satisfies $\mathcal{F}(0)=\dot{\mathcal{F}}(0)=0$ and $\lim_{\tau\to\infty}\mathcal{F}(\tau)e^{-\tau}=a_\varepsilon$, with $a_\varepsilon$ a positive constant. In the common case that $\alpha_1$ has no explicit time-dependence, $\mathcal{F}(\tau)=e^\tau+e^{-\tau}-2$. Note this is a singular perturbation solution, meaning it approximates the exact solution only locally in a specific region of phase space. This particular solution is no longer valid when $\mu\to 0$ (Fig.~\ref{fig:asyregions}). It is equivalent to the ``early time solutions'' explored in~\cite{Knowles2009,Cohen2013,Meisl2014} and several other studies.
	
Normally, when a perturbative solution is singular, the next step is to employ a ``singular perturbation technique'', of which there are several, to convert it into a uniformly convergent solution. However, all such techniques hitherto developed fail to regularize the singular perturbation solution Eqs.~\eqref{genseries1} and its higher order extensions. This is because protein aggregation kinetics is a very unusual kind of singular perturbation problem, as discussed extensively in~\cite{Dear2023B}: the region of validity in phase space of this perturbative solution can never be moved from the $\mu\to 1$ limit because the initial conditions themselves are being perturbed. This region of validity characterizes the $\mu\to 1$ asymptotic regime for the overall dynamics.

\subsection{$\mu\to 0$ asymptotic limit of dynamics}\label{sec:asylims}

We next explore the opposite limit of the dynamics: the $\mu\to 0$ or high-seed limit. We will make the common approximation that $m_c=0$. According to these equations, $\mu$ then decreases monotonically with $\tau$, reaching $\mu=0$ as $\tau\to\infty$. Conversely, $\Pi$ increases monotonically with $\tau$. Dividing Eq.~\eqref{geneqnondimP} by Eq.~\eqref{geneqnondimm} and integrating once yields:
	\begin{multline}\label{Pimugen}
		\Pi(\mu)=\left(p^2+4\varepsilon\frac{\alpha_e(m_\text{tot})}{\alpha_1(0,m_\text{tot})}\int_\mu^{\mu_0}\frac{\alpha_1(t(\mu),m_\text{tot}\mu)}{\alpha_e(m_\text{tot}\mu)}d\mu\right.\\\left.+2\frac{\alpha_e(m_\text{tot})}{\alpha_2(m_\text{tot})}\int_\mu^{\mu_0}\frac{\alpha_2(m_\text{tot}\mu)}{\alpha_e(m_\text{tot}\mu)}(1-\mu)d\mu\right)^{1/2}.
	\end{multline}
(To perform the integral explicitly when $\alpha_1$ has explicit $t$-dependence, $t(\mu)$ must be approximated, e.g.\ by inverting an approximate solution for $\mu$.) The kinetics can thus equivalently be described by a single differential equation:
\begin{equation}\label{mueqred}
\frac{d\mu}{d\tau}=-\frac{\alpha_e(m_\text{tot}\mu)}{\alpha_e(m_\text{tot})}\Pi(\mu)
\end{equation}
Provided $\lim_{\mu\to 0}\alpha_2/\alpha_e$ is finite, $\Pi$ will attain a maximal value $\Pi_\infty=\Pi(\tau\to\infty)$; otherwise, $\Pi_\infty$ will diverge to infinity. Either way, the structure in phase space of the dynamics of protein aggregation are relatively simple, featuring an attractive fixed point at $(\mu(\tau)=0,\ \Pi(\tau)=\Pi_\infty)$. The dynamics linearize around fixed points; here, they are found by expanding Eq.~\eqref{mueqred} in $\mu$, yielding:
\begin{equation}\label{mueqred3}
\frac{d\mu}{d\tau}=-\frac{2k_+m_\text{tot}}{\alpha_e(m_\text{tot})}\mu\Pi_\infty+O(\mu^2).
\end{equation}
Eq.~\eqref{mueqred3} is solved by (Fig.~\ref{fig:asyregions}):
\begin{equation}\label{mu3soln}
\mu(\tau)=A\exp\left(-\frac{2k_+m_\text{tot}}{\alpha_e(m_\text{tot})}\Pi_\infty\tau\right),
\end{equation}
where $A$ is an unknown constant of integration that is determined later.

This linearized region defines the $(\mu\to 0, \Pi\to \Pi_\infty)$ asymptotic regime for the dynamics; it is attained once $1-\Pi(\mu)/\Pi_\infty\ll 1$ and $\alpha_e\to 2k_+m_\text{tot}\mu$. If this occurs at large enough values of $\mu$ then this regime is important for the overall dynamics of $\mu$. (If $\Pi_\infty$ diverges, on the other hand, this regime is vanishingly small and the dynamics within it are irrelevant to the overall dynamics.)

Under high seeding conditions, the $\mu\to 0$ asymptotic limit is often entered much sooner (although this also depends on the precise form of the rate laws) and the corresponding region of phase space captures a much larger fraction of the overall dynamical trajectory, as $\Pi(\mu)/\Pi_\infty$ starts from a larger value. Moreover, inspection of Eq.~\eqref{Pimugen} reveals that $\Pi_\infty$, and therefore the $\mu\to 0$ dynamics, depends explicitly on the seed concentrations $\delta$ and $p$. Therefore, sufficiently high seeding levels affect the overall dynamics not just by modifying the dynamics within the $\mu\to 1$ asymptotic limit but also by modifying the $\mu\to 0$ dynamics. Comparison with Eq.~\eqref{genseries1} makes it clear that the functional dependence of the dynamics on $\delta$ and $p$ is likely to be very different in these two limits for the majority of plausible rate laws.

Note that since Eq.~\eqref{mueqred} is autonomous, we have reduced the problem to quadrature:
\begin{equation}
\tau=-\int_{1-\delta}^{\mu}\frac{\alpha_e(m_\text{tot})ds}{\alpha_e(m_\text{tot}s)\Pi(s)}.
\end{equation}
Although this technically qualifies as an exact solution of the differential equations, this integral cannot be performed explicitly, nor inverted, even for the simplest possible reactions (i.e.\ unsaturated, uninhibited, single-species kinetics). So, an analytical expression for $\mu$ and therefore $M$ remains accessible only by approximate techniques.
	
\subsection{Non-technical overview of derivation of global solutions}\label{sec:nontechoverview}

In~\cite{Dear2023B}, a mathematical approach based on Lie group theory referred to as the ``method of asymptotic symmetries'' was developed and employed to solve Eqs.~\eqref{MP1} when initial fibril (``seed'') concentrations are low or zero. Loosely speaking, ``asymptotic symmetries'' of an object are symmetry properties possessed by the object only in certain regions of phase space. It is convenient to label these symmetries according to the region of phase space in which they become asymptotically exact. For instance, $\mu\to 1$ asymptotic symmetries of Eqs.~\eqref{MP1} become exact symmetry properties only in the region of phase space defined by the limit $\mu\to 1$. In this study it is not necessary to understand the mathematical structure of these symmetries, only that they describe transformations of the parameters and variables upon which the object depends that leave the object unchanged, analogously to rotating a circle about its center. 

Under certain circumstances, asymptotic symmetries of an object become approximately valid globally, not just in the region of phase space in which they are defined. When this object is the solution to a differential equation, such globally valid asymptotic symmetries can then be exploited to transform special solutions like Eq.~\eqref{refsoln}, valid for specific parameter values, into global approximate solutions valid for all parameter values. A highly simplified analogy for this procedure using the example of a circle is finding a special solution to the equation for a circle valid at a specific value of the angle in polar co-ordinates, (i.e.\ finding a single point on its circumference); identifying an arbitrary rotation as a globally valid symmetry for the general solution; and transforming the special solution into the general one by applying this symmetry. Note this example is exact rather than approximate, however. In Appendix~\ref{app:transform} we describe these types of asymptotic symmetry and outline how the transformation procedure can be implemented mathematically; see refs.~\cite{Dear2025PRSA}-\cite{Dear2023B} for a more complete description.

For Eqs.~\eqref{MP1}, the circumstances under which $\mu\to 1$ asymptotic symmetries are approximately valid globally were found to be quite broad. However, for two reasons the specific asymptotic symmetry transformations used in~\cite{Dear2023B} can no longer be employed when seed concentrations are moderate or high. The first reason is that they no longer preserve the initial conditions of the rate equations, Eqs.~\eqref{ICs}. In Appendix~\ref{app:preserve} we demonstrate that this is a technical problem that can easily be overcome, and derive new $\mu\to 1$ asymptotic symmetries that do preserve initial conditions. 

The second reason is that the $\mu\to 1$ asymptotic symmetries are no longer always globally valid, depending on the precise forms of the rates. Essentially, this is because the effect of $p$ and $\delta$ on the $\mu\to 0$ dynamics becomes important for the overall dynamics and can no longer be ignored at higher seed concentrations, as discussed above. In Appendix~\ref{app:validitymu1} we give a fuller account of the conditions for global validity of $\mu\to 1$ asymptotic symmetries. To adequately model kinetics where the $\mu\to 0$ asymptotic region is important nonetheless still requires a uniformly valid symmetry with which to transform the special solution. We therefore propose a different type of asymptotic symmetry that we term a ``composite symmetry'', created by merging asymptotic symmetries from the $\mu\to 1$ and $\mu\to 0$ regions of phase space. The condition that must be satisfied for a composite symmetry to be uniformly valid throughout phase space (and why it is satisfied here) is presented in Appendix~\ref{app:composite}, alongside its application to transform the special solution into a global one.

The end result of these calculations is the following general solution, valid for any seed concentrations:
\begin{subequations}\label{gensoln}
	\begin{align}
	m(t)&=m(0)\left(1+\frac{p_0}{2c}\frac{\kappa_0}{\kappa}(e^{\kappa t}-e^{-\kappa t})\vphantom{\left(\frac{\varepsilon_0}{c}+\frac{\delta_0}{2c}+\frac{p_0^2}{2c}Q\right)}\right.\nonumber\\
	&\qquad\qquad+\left.\frac{\kappa_0^2}{\kappa^2}\left(\frac{\varepsilon_0}{c}+\frac{\delta_0}{2c}+\frac{p_0^2}{2c}Q\right)\mathcal{F}(\tau)\right)^{-c},\\
	\kappa_0&=\sqrt{\alpha_e(m(0))\alpha_2(m(0))},\qquad\delta_0=\frac{M(0)}{m(0)}, \\\varepsilon_0&=\frac{\alpha_1(m(0))}{2m(0)\alpha_2(m(0))},\qquad p_0=\frac{\alpha_e(m(0))}{m(0)\kappa_0}P(0) .
	\end{align}
\end{subequations}
Note, $\varepsilon_0$ and $\kappa_0$ recover $\varepsilon$ and $\kappa$ in the unseeded limit $M(0)\to 0$. The term proportional $Q$ is second-order in normalized seed concentration $p_0$ so can to a very good approximation be neglected when seed concentrations are not large. Nonetheless, if needed, it is given by:
\begin{equation}\label{Q}
Q=\frac{1+c}{c}-\frac{d\ln\alpha_e(m)}{d\ln m}\bigg|_{m=m(0)}.
\end{equation}
When the effect of the perturbation parameters including $\varepsilon,\ \delta$ and $p$ on the $\mu\to 0$ dynamics is unimportant for the global dynamics, $c$ is still given by the $\mu\to 1$ asymptotic symmetry transformation calculated in~\cite{Dear2023B}:
\begin{equation}\label{cmu1}
c=\frac{3}{2n_2'+1},\quad n_2'=\frac{d\ln\!\!\left[\alpha_2(m)\alpha_e(m)^2\right]}{d\ln m}\bigg|_{m=m_\text{tot}}\!\!\!\!\!\!\!\!\!\!\!\!-2.
\end{equation}
When this effect is instead important, $c$ is instead given by the $\mu\to 0$ asymptotic symmetry transformation calculated in Appendix~\ref{app:composite}:
\begin{equation}\label{cmu0}
c=\frac{2k_+m_\text{tot}}{\alpha_e(m_\text{tot})}\Pi_\infty.
\end{equation}
The determination of which formula to use for $c$ must be made on a case-by-case basis depending on the form of the rates $\alpha_1,\ \alpha_e$ and $\alpha_2$. It is also conceivable that for some rate equations where the $\mu\to 0$ asymptotic region is non-vanishing, the integral within $\Pi_\infty$ cannot be performed explicitly. In such cases, Eq.~\eqref{cmu1} could be used for $c$, although such a solution would lose accuracy at high seed concentrations. If so, it might be more appropriate to instead use $\Pi_0$ as an approximation for $\Pi_\infty$, and set $c=\Pi_0$.

\section{Experimental Methods}
\subsection{Materials}
The buffer chemicals were of analytical grade and Milli-Q water was used to prepare all buffer solutions.

\subsection{Recombinant A\textbeta40 and A\textbeta42 expression and purification}
A\textbeta(1-40) peptide, DAEFRHDSGYEVHHQKLVFFAEDVGSNKGAIIGLMVGGVV, and A\textbeta(M1-42) peptide, MDAEFRHDSGYEVHHQKLVFFAEDVGSNKGAIIGLMVGGVVIA were expressed in \emph{E.coli}. A\textbeta(M1-42) was purified using a combination of sonication, centrifugation, anion-exchange chromatography (AEX), and size exclusion chromatography (SEC), as described before~\cite{Linse2020b,Thacker2023}. A\textbeta(1-40) was expressed in fusion with the Npro autoprotease mutant EDDIE and pre-cleavage purified using sonication, centrifugation, AEX, cleaved in 1 M Tris/HCl, 5 mM DTT, 1 mM EDTA, pH 7.8 and post-cleavage purified using dialysis, AEX, and SEC as described previously~\cite{Linse2020}. The purified monomeric A\textbeta42 and A\textbeta40 were aliquoted, lyophilized and stored at -20 \celsius\ until further usage.

\subsection{Preparation of A\textbeta40 and A\textbeta42 seeds}

For the final isolation of monomeric peptide for the experiments, an aliquot was dissolved in 6 M GuHCl at 8.5 and subjected to SEC on a 1=7300 Superdex column in 20 mM NaP, 0.2 mM EDTA, 0.02\% \ce{NaN3} at pH 7.4 for A\textbeta40 and A\textbeta42. The freshly purified A\textbeta40 or A\textbeta42 were kept on ice before being aliquoted into half-area 96-well plates with PEGylated polystyrene surface (Corning 3881). The plates were placed at 37 \celsius\ in a plate reader (FluoStar Omega) in the presence of 1 \textmu M ThT. The ThT fluorescence intensities were measured through the bottom of the plate (continuous reading) using an excitation filter at 448 nm and an emission filter at 488 nm. The A\textbeta40 or A\textbeta42 fibrils were collected shortly after the kinetic curves had reached a plateau, hereafter referred to as seeds. 

The A\textbeta40 or A\textbeta42 seeds were then either kept idle at 37 \celsius\ in an incubator or treated with one of the following methods: tip sonication, freeze-thaw cycles or shaking. For tip sonication, solutions containing A\textbeta40 or A\textbeta42 seeds were kept on ice, and tip sonicated with 15 pulses (1s on/1s off) 10 times with a rest of around 1 minute in between using a microtip and a Soniprep 150 plus sonicator (MSE). For the freeze-thaw procedure, solutions were immediately frozen by dipping the tube containing A\textbeta40 or A\textbeta42 seeds into ethanol with dry ice, followed by thawing in 37 \celsius\ water. This freeze-thaw process was repeated 10 times with vortexing in between. Shaking was performed at 2000 rpm for 1 hour at 37 \celsius  with the sample in a eppendorf tube on a Eppendorf ThermoMixer C shaker . 
	
\subsection{Aggregation kinetics by ThT fluorescence}
The freshly purified proteins were distributed in the wells of half-area 96-well plates with PEGylated polystyrene surface (Corning 3881) with different concentrations of seeds. The aggregation of  A\textbeta40 or  A\textbeta42 monomers was monitored in the presence of 10 $\mu$M ThT at 37 \celsius\ through the bottom of the plate (continuous reading) using the same setup as for preparation of the A\textbeta40 and A\textbeta42 seeds. 

\subsection{Cryogenic Electron Microscopy (cryo-TEM)}
The solutions with seeds were pipetted on glow-discharged lacey formvar carbon-coated copper grids (Ted Pella), followed by blotting and plunging into liquid ethane at -184 \celsius. In this way, the specimens form thin liquid films in a glass-like state, avoiding the formation of ice crystals. The acceleration voltage of the electron microscope was 200 kV and zero-loss images were recorded digitally with a TVIPS F416 camera using SerialEM under low dose conditions with a 10 eV energy selecting slit in place. 

\begin{acknowledgments}
	We acknowledge support from the Lindemann Trust Fellowship, English-Speaking Union (AJD), the Swedish Research Council (SL) and NovoNordiskFonden (SL). The research leading to these results has received funding from the European Research Council under the European Union's Seventh Framework Programme (FP7/2007-2013) through the ERC grants PhysProt (agreement no. 337969) (TPJK, SL) and MAMBA (agreement no. 340890) (SL). We also acknowledge support from the Wellcome Trust (TPJK), the Cambridge Centre for Misfolding Diseases (TPJK), the BBSRC (TPJK), and the Frances and Augustus Newman foundation (TPJK).
\end{acknowledgments}
	
\appendix

\begin{widetext}
	
\section{Exact solution for fibril number concentration}\label{sec:exactPi}

In the case of saturated aggregation, the rates are:
\begin{equation}
\alpha_e(m)=\frac{2k_+m}{1+m/K_E},\qquad
\alpha_1(t,m)=\frac{k_nm^{n_c}}{1+(m/K_P)^{n_c}},\qquad
\alpha_2(m)=\frac{k_2m^{n_2}}{1+(m/K_S)^{n_2}}.
\end{equation}
Writing $\mathcal{K}_X$ for nondimensionalized saturation concentrations $K_X/m_\text{tot}$, the integral within the expression Eq.~\eqref{Pimugen} for $\Pi$ can be performed explicitly, yielding:
	\begin{subequations}
		\begin{multline}\label{Pimu}
		\Pi(\mu)=\left(p^2+4\varepsilon\frac{1+1/\mathcal{K}_P^{n_c}}{1+1/\mathcal{K}_E}\left(\frac{\mathcal{K}_P^{n_c}}{n_c}\ln\!\left[\frac{1+(\mu_0/\mathcal{K}_P)^{n_c}}{1+(\mu/\mathcal{K}_P)^{n_c}}\right]+\frac{Q_1(\mu,\mu_0,n_c,\mathcal{K}_P)}{\mathcal{K}_E}\right)\right.\\\left.+2\frac{1+1/\mathcal{K}_S^{n_2}}{1+1/\mathcal{K}_E}\left(\frac{\mathcal{K}_S^{n_2}}{n_2}\ln\!\left[\frac{1+(\mu_0/\mathcal{K}_S)^{n_2}}{1+(\mu/\mathcal{K}_S)^{n_2}}\right]-\left(1-\frac{1}{\mathcal{K}_E}\right)Q_{2a}(\mu,\mu_0,n_2,\mathcal{K}_S)-\frac{Q_{2b}(\mu,\mu_0,n_2,\mathcal{K}_S)}{\mathcal{K}_E}\right)\right)^{1/2},
		\end{multline}
		\begin{align}
		Q_1(\mu,\mu_0,n_c,\mathcal{K}_P)&=\frac{\mu_0^{1+n_c}}{1+n_c}{}_2\mspace{-2mu}F_1\!\!\left[1,1+\frac{1}{n_c},2+\frac{1}{n_c},-\left(\frac{\mu_0}{\mathcal{K}_P}\right)^{n_c}\right]-\frac{\mu^{1+n_c}}{1+n_c}\,{}_2\mspace{-2mu}F_1\!\!\left[1,1+\frac{1}{n_c},2+\frac{1}{n_c},-\left(\frac{\mu}{\mathcal{K}_P}\right)^{n_c}\right]\\
		Q_{2a}(\mu,\mu_0,n_2,\mathcal{K}_S)&=\frac{\mu_0^{1+n_2}}{1+n_2}{}_2\mspace{-2mu}F_1\!\!\left[1,1+\frac{1}{n_2},2+\frac{1}{n_2},-\left(\frac{\mu_0}{\mathcal{K}_S}\right)^{n_2}\right]-\frac{\mu^{1+n_2}}{1+n_2}\,{}_2\mspace{-2mu}F_1\!\!\left[1,1+\frac{1}{n_2},2+\frac{1}{n_2},-\left(\frac{\mu}{\mathcal{K}_S}\right)^{n_2}\right]\\
		Q_{2b}(\mu,\mu_0,n_2,\mathcal{K}_S)&=\frac{\mu_0^{2+n_2}}{2+n_2}{}_2\mspace{-2mu}F_1\!\!\left[1,1+\frac{2}{n_2},2+\frac{2}{n_2},-\left(\frac{\mu_0}{\mathcal{K}_S}\right)^{n_2}\right]-\frac{\mu^{2+n_2}}{2+n_2}\,{}_2\mspace{-2mu}F_1\!\!\left[1,1+\frac{2}{n_2},2+\frac{2}{n_2},-\left(\frac{\mu}{\mathcal{K}_S}\right)^{n_2}\right].
		\end{align}
	\end{subequations}
	${}_2\mspace{-2mu}F_1\!\!\left[a,b,c,z\right]$ is the Gauss hypergeometric function, and has the small-$z$ expansion: ${}_2\mspace{-2mu}F_1\!\!\left[a,b,c,z\right]=1+z\cdot ab/c+O(z^2)$. Therefore, if there is no saturation at all, i.e.\ $1/\mathcal{K}_X\to 0$, Eq.~\eqref{Pimu} simplifies to:
	\begin{equation}
	\Pi(\mu)=\left(p^2+\frac{4\varepsilon}{n_c}\left(\mu_0^{n_c}-\mu^{n_c}\right)+2\left(\frac{\mu_0^{n_2}-\mu^{n_2}}{n_2}-\frac{\mu_0^{n_2+1}-\mu^{n_2+1}}{n_2+1}\right)\right)^{1/2}.\label{Piunsat}
	\end{equation}

Eq.~\eqref{Pimu} also permits $\Pi_\infty$ to be calculated as $\Pi(\mu=0)$. In reality $\Pi_\infty=\Pi(\mu=\mu_s)$, where $\mu_s=m_s/m_\text{tot}$ is the nondimensionalized solubility limit, but $\mu_s\ll 1$. Provided also $K_X \gg m_s$, as it generally is, the approximation $\Pi_\infty=\Pi(\mu=0)$ is a good one, and yields:
\begin{multline}\label{Piinfty}
	\Pi_\infty=\left(p^2+4\varepsilon\frac{1+1/\mathcal{K}_P^{n_c}}{1+1/\mathcal{K}_E}\left(\frac{\mathcal{K}_P^{n_c}}{n_c}\ln\!\left[1+(\mu_0/\mathcal{K}_P)^{n_c}\right]+\frac{1}{\mathcal{K}_E}\cdot\frac{\mu_0^{1+n_c}}{1+n_c}{}_2\mspace{-2mu}F_1\!\!\left[1,1+\frac{1}{n_c},2+\frac{1}{n_c},-\left(\frac{\mu_0}{\mathcal{K}_P}\right)^{n_c}\right]\right)\right.\\\left.+2\frac{1+1/\mathcal{K}_S^{n_2}}{1+1/\mathcal{K}_E}\left(\frac{\mathcal{K}_S^{n_2}}{n_2}\ln\!\left[1+(\mu_0/\mathcal{K}_S)^{n_2}\right]-\left(1-\frac{1}{\mathcal{K}_E}\right)Q_{2a}(0,\mu_0,n_2,\mathcal{K}_S)-\frac{Q_{2b}(0,\mu_0,n_2,\mathcal{K}_S)}{\mathcal{K}_E}\right)\right)^{1/2}.
\end{multline}
(When fragmentation is present, i.e.\ $n_2=0$, the integration yields a $\ln\mu$ term, and $\Pi(\mu)$ no longer reaches a constant value as $\mu\to 0$. This aberrant behaviour occurs because of the approximations made that the fragmentation rate at a given site does not depend on the length of the parent fibril, and the neglection of full mass balance. Although valid on the aggregation timescale,  these approximations are not valid as $\tau\to\infty$, and in practice a constant $\Pi_\infty$ will be achieved. However, this value is relevant only long after the aggregation has completed, so its calculation is not particularly useful.)
	
\section{Calculating $P_\infty$ when secondary nucleation is unsaturated}\label{app:amylofit}

When secondary nucleation is unsaturated, $\mathcal{K}_S\to \infty$ and $\Pi_\infty$ reduces to:
		\begin{multline}\label{PiinftynoKS}
		\Pi_\infty=\left(p^2+4\varepsilon\frac{1+1/\mathcal{K}_P^{n_c}}{1+1/\mathcal{K}_E}\left(\frac{\mathcal{K}_P^{n_c}}{n_c}\ln\!\left[1+(\mu_0/\mathcal{K}_P)^{n_c}\right]+\frac{1}{\mathcal{K}_E}\cdot\frac{\mu_0^{1+n_c}}{1+n_c}{}_2\mspace{-2mu}F_1\!\!\left[1,1+\frac{1}{n_c},2+\frac{1}{n_c},-\left(\frac{\mu_0}{\mathcal{K}_P}\right)^{n_c}\right]\right)\right.\\\left.+2\frac{1}{1+1/\mathcal{K}_E}\left(\frac{\mu_0^{n_2}}{n_2}-\frac{\mu_0^{1+n_2}}{1+n_2}+\frac{1}{\mathcal{K}_E}\left(\frac{\mu_0^{1+n_2}}{1+n_2}-\frac{\mu_0^{2+n_2}}{2+n_2}\right)\right)\right)^{1/2}.
		\end{multline}
		
When elongation is also unsaturated, it reduces to:
		\begin{equation}\label{PiinftynoKSKE}
		\Pi_\infty=\left(p^2+\frac{4\varepsilon}{n_c} (1+\mathcal{K}_P^{n_c})\ln\!\left[1+(\mu_0/\mathcal{K}_P)^{n_c}\right]+2\left(\frac{\mu_0^{n_2}}{n_2}-\frac{\mu_0^{1+n_2}}{1+n_2}\right)\right)^{1/2}.
		\end{equation}
	
\section{Calculating $P_\infty$ when secondary nucleation is saturated}\label{app:avelength}
	
	In the limit that only secondary nucleation is saturated, the final argument of the hypergeometric functions in $Q_{2a}(0,\mu_0,n_2,\mathcal{K}_S)$ and $Q_{2b}(0,\mu_0,n_2,\mathcal{K}_S)$ blows up. Two identities help us. First:
	\begin{subequations}
		\begin{align}
		&{}_2\mspace{-2mu}F_1\!\!\left[a,b,c,z\right]=(1-z)^{-a}{}_2\mspace{-2mu}F_1\!\!\left[a,c-b,c,\frac{z}{z-1}\right]\\
		\therefore &\lim_{z\to-\infty}{}_2\mspace{-2mu}F_1\!\!\left[1,1+x,2+x,z\right]=-\frac{1}{z}{}_2\mspace{-2mu}F_1\!\!\left[1,1,2+x,1\right].
		\end{align}
	\end{subequations}
	Next:
	\begin{subequations}
		\begin{align}
		&{}_2\mspace{-2mu}F_1\!\!\left[a,b,c,1\right]=\frac{\Gamma(c)\Gamma(c-a-b)}{\Gamma(c-a)\Gamma(c-b)},\ \text{Re} (c-a-b)>0\\
		\therefore&{}_2\mspace{-2mu}F_1\!\!\left[1,1,2+x,1\right]=\frac{\Gamma(2+x)\Gamma(x)}{\Gamma(1+x)^2}=\frac{(1+x)\Gamma(1+x)\Gamma(x)}{x\Gamma(1+x)\Gamma(x)}=1+\frac{1}{x}.
		\end{align}
	\end{subequations}
	
	Combining these results:
		\begin{subequations}
			\begin{align}
			\lim_{\mathcal{K}_S\to 0}Q_{2a}(0,\mu_0,n_2,\mathcal{K}_S)&=\lim_{\mathcal{K}_S\to 0}\frac{\mu_0^{1+n_2}}{1+n_2}{}_2\mspace{-2mu}F_1\!\!\left[1,1+\frac{1}{n_2},2+\frac{1}{n_2},-\left(\frac{\mu_0}{\mathcal{K}_S}\right)^{n_2}\right]\\
			&\qquad=\frac{\mu_0\mathcal{K}_S^{n_2}}{1+n_2}\cdot(1+n_2)=\mu_0\mathcal{K}_S^{n_2} \label{limQ2asatsec}\\
			\lim_{\mathcal{K}_S\to 0}Q_{2b}(0,\mu_0,n_2,\mathcal{K}_S)&=\lim_{\mathcal{K}_S\to 0}\frac{\mu_0^{2+n_2}}{2+n_2}{}_2\mspace{-2mu}F_1\!\!\left[1,1+\frac{2}{n_2},2+\frac{2}{n_2},-\left(\frac{\mu_0}{\mathcal{K}_S}\right)^{n_2}\right]\\
			&\qquad=\frac{\mu_0^{2}\mathcal{K}_S^{n_2}}{2+n_2}\cdot\left(1+\frac{n_2}{2}\right)=\frac{\mu_0^2}{2}\mathcal{K}_S^{n_2}.
			\end{align}
		\end{subequations}
		So, assuming that $\varepsilon$ is small enough that it may be neglected to leading order, $\Pi_\infty$ becomes:
		\begin{equation}
		\text{So,   }\Pi_\infty\simeq\left(p^2+\frac{2}{\mathcal{K}_S^{n_2}}\left(\frac{\mathcal{K}_S^{n_2}}{n_2}\ln\!\left[1+(\mu_0/\mathcal{K}_S)^{n_2}\right]-Q_{2a}(0,\mu_0,n_2,\mathcal{K}_S)\right)\right)^{1/2}\simeq \left(p^2+2\ln\!\left[\mu_0/\mathcal{K}_S\right]-2\mu_0\right)^{1/2}.
		\end{equation}
		
		In reality, as discussed above, $\Pi_\infty=\Pi(\mu=\mu_s)$, not $\Pi(\mu=0)$. From Eq.~\eqref{limQ2asatsec}, the limit $\lim_{\mathcal{K}_S\to 0}Q_{2a}(\mu_s,\mu_0,n_2,\mathcal{K}_S)$ can easily enough be seen to be $(\mu_0-\mu_s)\mathcal{K}_S^{n_2}$. Using Eq.~\eqref{Pimu}, and with $1/\mathcal{K}_E\to 0$, $\Pi(\mu_s)$ thus becomes:
		\begin{equation}
		\Pi_\infty(K_S\ll m_s)=\lim_{\mathcal{K}_S\to 0}\Pi(\mu_s)=\left(p^2+2\ln\!\left[\mu_0/\mu_s\right]-2(\mu_0-\mu_s)\right)^{1/2}.\label{Pisol}
		\end{equation}

\end{widetext}	
	
	\section{Derivation of required seed concentration to bypass all nucleation}\label{app:M0crit}
	
	We define the critical seed concentration $M_\text{crit,2}$ as follows: the number concentration of seeds, $P_0$, is equivalent to the number concentration of new aggregates formed over the course of the reaction. To obtain a rough estimate of the number of new aggregates formed over the course of the reaction (in the absence of elongation saturation), we make two assumptions. First, we estimate the time taken to convert a fraction $x$ of the monomer present into fibrils to obtain an estimate for the time to completion. As we use this to determine the time taken to finish the experiment when seed elongation dominates the reaction, we can self-consistently assume that the increase in mass is mainly due to the elongation of seeds, thus obtaining a simplified moment equation for M:
	\begin{equation}
	\frac{dM}{dt}= 2 k_+ m(t) P_0
	\end{equation}
	which can easily be solved to yield
	\begin{equation}
	M(t)=M_0+m_0(1-\exp(-2 k_+P_0t))
	\end{equation}
	and therefore the time to convert a fraction $x$, of the monomer present, $m_0$, is given by
	\begin{equation}
	t_x= -\frac{1}{2 k_+P_0} \ln (1-x).
	\end{equation}
	For convenience, we set $\ln(1-x)=-1$ which is equivalent to approximately $2/3$ conversion and yields
	\begin{equation}
	t_{\mathrm{end}}= \frac{1}{2 k_+P_0}.
	\label{eq:t_end_simple}
	\end{equation}
	
	Next, to estimate the number of new aggregates formed over the course of $t_{\mathrm{end}}$ we only consider the seed fibrils as sources of new aggregates, and linearize the equations by $m(t)\to m_0$, giving  
	\begin{equation}
	\frac{dP}{dt}= \alpha_2(m_0)(M_0+ m_0(1-\exp(-2 k_+P_0t))),
	\end{equation}
	where as usual $\alpha_2(m)M$ is the rate of secondary nucleation. With boundary condition $P(0)=P_0$, the solution yields
	\begin{equation}
	P(t)= P_0 + \alpha_2(m_0)\left(M_0t+ m_0\left(t+\frac{e^{-2 k_+P_0t}-1}{2k_+P_0}\right)\right).
	\end{equation}
	Using the above derived $t_{\mathrm{end}}=\frac{1}{2 k_+P_0}$, we get
	\begin{multline}
	P(t_\mathrm{end})= P_0 + \frac{\alpha_2(m_0)}{2k_+P_0}(M_0+ \frac{m_0}{e})\\=P_0\left(1 + \frac{\alpha_2(m_0)}{2k_+}\frac{M_0}{P_0^2}\left(1+\frac{1}{e}\frac{m_0}{M_0}\right)\right)\\=P_0\left(1 + \frac{L_0^2}{L_{ss0}^2}\frac{m_0}{M_0}\left(1+\frac{1}{e}\frac{m_0}{M_0}\right)\right),
	\end{multline}
	where $L_{ss0}$ is the steady-state fibril length that would be achieved in this reaction if it were performed without seeds, and is given by:
    \begin{equation}
        L_{ss0}=\sqrt{\frac{2k_+m_0}{\alpha_2(m_0)}}.
    \end{equation}
    Setting $P(t_\mathrm{end})$ equal to $2P_0$, i.e. a doubling of the initial number of aggregates, gives:
	\begin{equation}
	\frac{L_0^2}{L_{ss0}^2}\frac{m_0}{M_0}\left(1+\frac{1}{e}\frac{m_0}{M_0}\right)=1.
	\end{equation}
	
\begin{figure}
		\centering
		\includegraphics[width=0.48\textwidth]{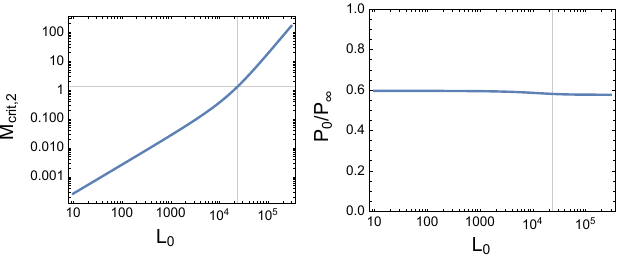}
		\caption{Behaviour of Eq.~\ref{equ:Mcrit2}. \textbf{a} Critical aggregate concentration $M_{crit,2}$, as a function of the initial seed length $L_0$. Note the linear regime at low $L_0$ and quadratic regime at high $L_0$. The grey lines correspond to $L_{ss0}$, where this transition occurs. \textbf{b} The ratio of $P_0 = M_{crit,2}/L_0$, from Eq.~\ref{equ:Mcrit2}, over $P_{\infty}$, evaluated by integration of the moment equations. The definition of $M_{crit,2}$ was that this ratio should be 0.5. Thus, the approximate expression for $M_{crit,2}$ from Eq.~\ref{equ:Mcrit2} performs well. Parameters as in Fig.~\ref{fig:Elon_only}.}
		\label{fig:SI_Mcrit_performance_figure}
	\end{figure}

    Solving for $M_0$ gives $M_\text{crit,2}$, the seed mass concentration at which the number of seeds is equivalent to the number of new aggregates formed over the course of the reaction (Eq.~\eqref{equ:Mcrit2}). Alternatively, if the amount of seed is given, one can solve for $L_0$ to obtain the seed length required instead. The required seed amounts, as a function of the initial seed length, are shown in Fig.~\ref{fig:SI_Mcrit_performance_figure}a. The performance of this approximation, evaluated against numerical integration of the moment equations is shown in Fig.~\ref{fig:SI_Mcrit_performance_figure}b, confirming that the above approximation matches well the definition of $M_{crit,2}$.
    
	In systems without appreciable secondary processes, there is no steady-state average fibril length. However, the average fibril length at the end of an unseeded reaction, $L_{\infty 0}$ can be computed from the known analytical solution~\cite{Oosawa1962,Cohen2011d}:
	\begin{equation}
	L_{\infty 0}^2=\frac{n_c}{2}\frac{2k_+m_0^2}{k_nm_0^{n_c}}.
	\end{equation}
	By its derivation, $t_\text{end}$ is unchanged by the absence of secondary processes, but the expression for $P(t)$ becomes:
	\begin{equation}
	P(t)=P_0+k_nm_0^{n_c}t.
	\end{equation}
	Setting $P(t_\text{end})=2P_0$, we have:
	\begin{align}
	P_0&=\frac{k_nm_0^{n_c}}{2k_+P_0}\\
	\Rightarrow \frac{M_0^2}{L_0^2}&=\frac{n_c}{2}\frac{m_0^2}{L_{\infty 0}^2}\\
	\Rightarrow \frac{M_0}{m_0}&=\sqrt{\frac{n_c}{2}}\frac{L_0}{L_{\infty 0}}.
	\end{align}
    
\begin{widetext}

\section{Method of asymptotic symmetries}\label{app:asymptoticmethod}

\subsection{Procedure for transformation of the special solution}\label{app:transform}
	
The full details of the original mathematical method are described in~\cite{Dear2023B}; in simple terms, it comprises 2 elements. First, identification of a special solution valid for certain values of the perturbation parameters. For Eqs.~\eqref{geneqsnondim} we have identified this to be Eq.~\eqref{refsoln}~\cite{Dear2023B}. Second, the identification of a continuous symmetry property of the dynamical equations connecting perturbation parameters to the parameters remaining in the special solution, and its use to transform the special solution into a global one valid for all parameter values. 

Eqs.~\eqref{geneqsnondim} unfortunately do not possess such symmetries valid uniformly throughout phase space, either exact or approximate. Instead,~\cite{Dear2023B} showed that so-called ``asymptotic symmetries'' of the (unknown) solution to Eqs.~\eqref{geneqsnondim}, that become exact only asymptotically in the $\mu\to 1$ asymptotic region, could be easily calculated from the perturbation series. It was also showed that these symmetries become approximately valid globally under a quite broad range of circumstances (that we outline in Appendix~\ref{app:validitymu1}), and could therefore be used to transform Eq.~\eqref{refsoln} into a general solution.

As explained in detail in~\cite{Dear2023B}, asymptotic symmetry transformations of the special solution Eq.~\eqref{refsoln} can in practice be performed without explicit computation of the asymptotic symmetries themselves. Instead, one simply replaces $c$ and $\delta$ in the special solution with functions $\tilde{c}$ and $\tilde{\delta}$ that ensure the asymptotic limit of the special solution matches that of the exact dynamics. The $\mu\to 1$ limit of the symmetry-transformed special solution Eq.~\eqref{refsoln} can be attained by premultiplying $\tilde{\delta}$ by $s$ and expanding in $s$ to second order:
\begin{equation}
\lim_{\mu\to 1}\mu^*(\tau,\tilde{c},\tilde{\delta})=1-s\tilde{\delta} e^{\tau}+s^2\frac{1+\tilde{c}}{2\tilde{c}}e^{2\tau}\tilde{\delta}^2+\mathcal{R},
\end{equation}
where $\mathcal{R}$ consists of terms of $O(\tilde{\delta}^2)$ that grow less rapidly in $\tau$ than $e^{2\tau}$. The $\mu\to 1$ limit of the dynamics to second order in $\delta,\ p$ and $\varepsilon$ is given by the perturbative solution Eq.~\eqref{genseries1} extended to second order:
\begin{equation}\label{genseries2}
\lim_{\mu\to 1}\mu=1-s\left[\vphantom{\frac{p}{2}}\varepsilon\mathcal{F}(\tau)+\frac{\delta}{2}(e^\tau+e^{-\tau})+\frac{p}{2}(e^\tau-e^{-\tau})\right]+\frac{s^2}{3}\left(a_\varepsilon\varepsilon+\frac{\delta}{2}+\frac{p}{2}\right)^2e^{2\tau}\left(\frac{d\ln\alpha_2}{d\ln m}+2\frac{d\ln\alpha_e}{d\ln m}\right)\!\!\bigg|_{m=m_\text{tot}}\!\!\!\!+\mathcal{R},
\end{equation}
where $a_\varepsilon$ is a positive constant satisfying $\lim_{\tau\to\infty}\mathcal{F}(\tau)e^{-\tau}=a_\varepsilon$ (see Sec.~\ref{sec:muto1} and~\cite{Dear2023B}). So, the second order symmetry transformation involves the following first-order transformation of $\delta$ in the special solution to $\tilde{\delta}$:
\begin{equation}\label{mu1transfOs}
e^\tau\tilde{\delta}(\delta, p, \varepsilon)=\varepsilon\mathcal{F}(\tau)+\frac{\delta}{2}(e^\tau+e^{-\tau})+\frac{p}{2}(e^\tau-e^{-\tau})+O((\delta,p,\varepsilon)^2).
\end{equation}
This specific transformation was used in~\cite{Dear2023B} to produce a general solution valid for $\varepsilon,\delta,p\ll 1$, i.e.\ low or zero seed concentrations, and secondary process-dominated kinetics. ($\tilde{c}$ is similarly given by the matching condition for the most-divergent terms in $\tau$ at second order in these expansions.) This solution fails for moderate seed concentrations primarily because it no longer satisfies the initial conditions. However, it is quite easily generalized to do so. We may in fact choose any $\tilde{\delta}$ whose first-order expansion satisfies Eq.~\eqref{mu1transfOs}, since all such values of $\tilde{\delta}$ can be arrived at by an $O(s)$ symmetry transformation. The general expression for $\tilde{\delta}$ is thus:
\begin{equation}\label{transf}
ce^{\tau}\left[(1-\tilde{\delta}(\delta, p, \varepsilon))^{-1/c}-1\right]=f_\delta(e^\tau+e^{-\tau})+f_\varepsilon\mathcal{F}(\tau)+f_p(e^\tau-e^{-\tau})+O((\delta,p,\varepsilon)^2),
\end{equation}
where $(f_\delta,f_p,f_\varepsilon)=(\delta/2,p/2,\varepsilon)+O((\delta,p,\varepsilon)^2)$. Remembering that the special solution Eq.~\eqref{refsoln} is:
\begin{subequations}\label{gengensoln}
	\begin{align}
	\mu^*(\tau,c,\delta)&=\frac{1}{\left(1+e^\tau\left[(1-\delta)^{-1/c}-1\right]\right)^{c}},\\
	c&=\frac{3}{2n_2+1},\qquad p^*=\sqrt{2\frac{1-(1-\delta)^{n_2}}{n_2}-2\frac{1-(1-\delta)^{n_2+1}}{n_2+1}},
	\end{align}
\end{subequations}
it is transformed by this more general transformation to:
\begin{equation}\label{gengensoln}
	\mu(\tau,c,\delta)=\left(1+\frac{f_\delta}{c}(e^\tau+e^{-\tau})+\frac{f_\varepsilon}{c}\mathcal{F}(\tau)+\frac{f_p}{c}(e^\tau-e^{-\tau})\right)^{-c}.
\end{equation}

\subsection{Determining a $\mu\to 1$ asymptotic symmetry transformation that preserves initial conditions}\label{app:preserve}
We may take advantage of this flexibility to choose $(f_\delta,f_p,f_\varepsilon)$ such that the transformed Eq.~\eqref{refsoln} also matches precisely the early-time kinetics up to $O(\tau^2)$. This choice of $\tilde{\delta}$ still satisfies Eq.~\eqref{mu1transfOs}. 

\subsubsection{Taylor series of $\mathcal{F}(\tau)$}
This procedure will first require knowledge of the expansion in $\tau$ of $\mathcal{F}$, which can be found by expanding in $\tau$ the first order perturbation equations governing the $\mu\to 1$ dynamics, Eqs.~\eqref{perteqsO1}, using $\mu^{(1)}=\sum_i\tau^i\mu^{(1)}_i$ and $\Pi^{(1)}=\sum_i\tau^i\Pi^{(1)}_i$:
\begin{align}
\frac{d\Pi^{(1)}}{d\tau}&=2\varepsilon\frac{\alpha_1(t,m_\text{tot})}{\alpha_1(0,m_\text{tot})}-\mu^{(1)}(\tau)\\
	\frac{d\mu^{(1)}}{d\tau}&=-\Pi^{(1)}(\tau)\\
	\Rightarrow\Pi^{(1)}_1&=2\varepsilon-\mu^{(1)}_0+O(\tau),\\
	\mu^{(1)}_1+2\tau\mu^{(1)}_2&=-\Pi^{(1)}_0-\tau\Pi^{(1)}_1+O(\tau^2).
\end{align}
$\Pi^{(1)}_0$ and $\mu^{(1)}_0$ are given by initial conditions as for the first order perturbation equation as $\Pi^{(1)}_0=p$ and $\mu^{(1)}_0=-\delta$. Matching order-by-order in $\tau$ gives:
\begin{equation}
\mu^{(1)}=-\delta-p\tau-\left(\varepsilon+\frac{\delta}{2}\right)\tau^2.
\end{equation}
Since the full $\mu^{(1)}$ is given by:
\begin{equation}
\mu^{(1)}=-\varepsilon\mathcal{F}(\tau)-\frac{\delta}{2}(e^\tau+e^{-\tau})-\frac{p}{2}(e^\tau-e^{-\tau}),
\end{equation}
we deduce that $\mathcal{F}(\tau)=\tau^2+O(\tau^3)$.

\subsubsection{Early-time expansion of the kinetics}
The early-time kinetics for $\Pi$ and $\mu$ can be solved fully from the unperturbed rate equations Eqs.~\eqref{geneqsnondim} to $O(\tau^2)$ in $\mu$ by substituting $\Pi=\Pi_0+\tau\Pi_1$ and $\mu=\mu_0+\tau\mu_1+\tau^2\mu_2$ into them. From the initital conditions, clearly $\Pi_0=sp$ and $\mu_0=1-s\delta$. Eq.~\eqref{geneqnondimP} consequently becomes:
\begin{equation}
\Pi_1=2s\varepsilon\frac{\alpha_1(0,m_0)}{\alpha_1(0,m_\text{tot})}+s\delta\frac{\alpha_2(m_0)}{\alpha_2(m_\text{tot})}.
\end{equation}
This can be simplified as follows:
\begin{align}
\Pi_1&=2s\frac{\alpha_1(0,m_\text{tot})}{2m_\text{tot}\alpha_2(m_\text{tot})}\frac{\alpha_1(0,m_0)}{\alpha_1(0,m_\text{tot})}+s\frac{M(0)}{m_\text{tot}}\frac{\alpha_2(m_0)}{\alpha_2(m_\text{tot})}\nonumber\\
&=\frac{m_0}{m_\text{tot}}\frac{\alpha_2(m_0)}{\alpha_2(m_\text{tot})}\left(2s\frac{\alpha_1(0,m(0))}{2m(0)\alpha_2(m(0))}+s\frac{M(0)}{m(0)}\right)\nonumber\\
\therefore \Pi_1&=\mu_0\frac{\alpha_2(m_0)}{\alpha_2(m_\text{tot})}\left(2s\varepsilon_0+s\delta_0\right),\quad \varepsilon_0=\frac{\alpha_1(0,m(0))}{2m(0)\alpha_2(m(0))}, \quad\delta_0=\frac{M(0)}{m(0)}.
\end{align}
Eq.~\eqref{geneqnondimm} becomes:
\begin{equation}
	\mu_1+2\tau\mu_2=-\frac{\alpha_e(m_0)}{\alpha_e(m_\text{tot})}\left(sp+\Pi_1\tau\right)-sp\cdot\tau\left(\frac{d}{d\tau}\frac{ \alpha_e(m)}{\alpha_e(m_\text{tot})}\right)\bigg|_{\tau=0}.
\end{equation}

We can write the final term as:
\begin{equation}
-sp\cdot\tau\left(\frac{d}{d\tau}\frac{ \alpha_e(m)}{\alpha_e(m_\text{tot})}\right)\bigg|_{\tau=0}=-sp\cdot\tau\left(\frac{dm}{d\tau}\frac{d}{dm}\frac{ \alpha_e(m)}{\alpha_e(m_\text{tot})}\right)\bigg|_{\tau=0}=-sp\cdot\tau \frac{m_\text{tot}}{\alpha_e(m_\text{tot})}\mu_1\frac{d\alpha_e}{dm}\bigg|_{m=m_0}.
\end{equation}
The terms at $O(\tau^0)$ are collected and simplified as follows:
\begin{subequations}
	\begin{align}
	\mu_1&=-sp\frac{\alpha_e(m_0)}{\alpha_e(m_\text{tot})}.\quad\text{Remembering }p=\frac{\alpha_e(m_\text{tot})}{m_\text{tot}\kappa}P(0),\ \text{this simplifies to}\ -s\frac{\alpha_e(m_0)}{m_\text{tot}\kappa}P(0).\nonumber\\
	\therefore \mu_1&=-sp_0\mu_0\frac{\kappa_0}{\kappa},\\
     p_0&=\frac{\alpha_e(m(0))}{m(0)\kappa_0}P(0),\quad
	\kappa_0=\sqrt{\alpha_e(m(0))\alpha_2(m(0))}.
	\end{align}
\end{subequations}
The terms at $O(\tau^1)$ are then collected and simplified as follows:
\begin{subequations}
	\begin{align}
	\mu_2&=-\frac{\alpha_e(m_0)}{\alpha_e(m_\text{tot})}\frac{\Pi_1}{2}-s\frac{p}{2} \frac{m_\text{tot}}{\alpha_e(m_\text{tot})}\mu_1\frac{d\alpha_e}{dm}\bigg|_{m=m_0}=-\frac{\alpha_e(m_0)}{\alpha_e(m_\text{tot})}\frac{\Pi_1}{2}-\frac{s\mu_1}{2\kappa} \frac{d\alpha_e}{dm}\bigg|_{m=m_0}\\
	&=-\frac{\mu_0}{2}\frac{\alpha_e(m_0)}{\alpha_e(m_\text{tot})}\frac{\alpha_2(m_0)}{\alpha_2(m_\text{tot})}\left(2s\varepsilon_0+s\delta_0\right)+sp_0\mu_0\frac{\kappa_0}{\kappa}\cdot \frac{s\alpha_e(m_0)}{2\kappa} \frac{d\ln\alpha_e}{dm}\bigg|_{m=m_0}\\
	&=-s\mu_0\frac{\kappa_0^2}{\kappa^2}\left(\varepsilon_0+\frac{\delta_0}{2}\right)+s^2p_0^2\mu_0\frac{\kappa_0^2}{\kappa^2}\frac{m(0)}{2}\frac{d\ln\alpha_e(m)}{dm}\bigg|_{m=m(0)}.
	\end{align}
\end{subequations}
In summary, we have:
\begin{subequations}
\begin{align}
\mu&=\mu_0+\tau\mu_1+\tau^2\mu_2+O(\tau^3)\\
\mu_0&=1-s\delta,\qquad\mu_1=-sp_0\mu_0\frac{\kappa_0}{\kappa},\\
\mu_2&=s\mu_0\frac{\kappa_0^2}{\kappa^2}\left(\varepsilon_0+\frac{\delta_0}{2}\right)+s^2p_0^2\mu_0\frac{\kappa_0^2}{\kappa^2}\frac{m(0)}{2}\frac{d\ln\alpha_e(m)}{dm}\bigg|_{m=m(0)}.
\end{align}
\end{subequations}

\subsubsection{Constraining the general solution on the early-time kinetics}
Next, imposing $\mu(0)=\mu_0=1-\delta$, we can eliminate one degree of freedom straightforwardly and rewrite the unconstrained general solution Eq.~\eqref{gengensoln} as:
\begin{equation}\label{gensolnseed}
    \mu=\mu_0\left(1+\frac{f_q}{2c}(e^{\tau}-e^{-\tau})+\frac{f_e}{c}\mathcal{F}(\tau)\right)^{-c},
\end{equation}
where $f_e$ and $f_q$ are now unknown functions that satisfy $f_e+f_q/2=s(\varepsilon+\delta/2+p/2)+O(s^2)$. Expanding this in $\tau$ gives:
\begin{equation}
    \mu=\mu_0-\mu_0 f_q\tau+\mu_0\left(f_q^2\frac{1+c}{2c}-f_e\right)\tau^2+O(\tau^3).
\end{equation}

Equating terms, we are left with:
\begin{subequations}
	\begin{align}
	f_q&=-\frac{\mu_1}{\mu_0}=sp_0\frac{\kappa_0}{\kappa}\\
	f_e&=f_q^2\frac{1+c}{2c}-\frac{\mu_2}{\mu_0}=s^2\frac{p_0^2}{2}\frac{\kappa_0^2}{\kappa^2}\left(\frac{1+c}{c}-\frac{d\ln\alpha_e(m)}{d\ln m}\bigg|_{m=m(0)}\right)+s\frac{\kappa_0^2}{\kappa^2}\left(\varepsilon_0+\frac{\delta_0}{2}\right).
	\end{align}
\end{subequations}

Pulling all this together, and setting $s=1$, we have the general solution:
\begin{subequations}
	\begin{align}
	m(t)&=m(0)\left(1+\frac{p_0}{2c}\frac{\kappa_0}{\kappa}(e^{\kappa t}-e^{-\kappa t})+\frac{\kappa_0^2}{\kappa^2}\left(\frac{\varepsilon_0}{c}+\frac{\delta_0}{2c}+\frac{p_0^2}{2c}Q\right)\mathcal{F}(\tau)\right)^{-c},\\
	\kappa&=\sqrt{\alpha_e(m_\text{tot})\alpha_2(m_\text{tot})},\qquad c=\frac{3}{2n_2'+1},\qquad n_2'=\frac{d\ln\!\!\left[\alpha_2(m)\alpha_e(m)^2\right]}{d\ln m}\bigg|_{m=m_\text{tot}}-2,\\
	\kappa_0&=\sqrt{\alpha_e(m(0))\alpha_2(m(0))},\qquad\varepsilon_0=\frac{\alpha_1(m(0))}{2m(0)\alpha_2(m(0))}, \qquad\delta_0=\frac{M(0)}{m(0)},\qquad p_0=\frac{\alpha_e(m(0))}{m(0)\kappa_0}P(0) .
	\end{align}
\end{subequations}
Note, $\varepsilon_0$ and $\kappa_0$ recover $\varepsilon$ and $\kappa$ in the unseeded limit $M(0)\to 0$. The term proportional $Q$ is second-order in normalized seed concentration $p_0$ so can to a very good approximation be neglected when seed concentrations are not large. Nonetheless, if needed, it is given by:
\begin{equation}\label{Q}
Q=\frac{1+c}{c}-\frac{d\ln\alpha_e(m)}{d\ln m}\bigg|_{m=m(0)}.
\end{equation}
	
\subsection{Global validity of $\mu\to 1$ asymptotic symmetries}\label{app:validitymu1}

The circumstances under which the $\mu\to 1$ asymptotic symmetry is approximately valid globally are as follows. First, if the $\mu\to 1$ asymptotic region encompasses most of the phase space, with the remainder having functionally similar behaviour to the special solution. Second, if all perturbation parameters and at least one non-perturbation parameter are primarily important only in this region, in which case the lack of accuracy outside this region of the asymptotic symmetry connecting them is immaterial. To these we add a third here: if the perturbation parameters are also important outside this region, but their relationship to the non-perturbation parameters here remains similar to within the region, in which case any errors following from using the symmetry here will be small.

\subsection{Solution by composite asymptotic symmetry}\label{app:composite}
	
If these conditions are violated, due e.g.\ to high seeding levels, an alternative type of symmetry is needed. If each non-perturbation parameter in the special solution comes to prominence in a different region of phase space, then the global symmetry properties of the problem can be captured by combining asymptotic symmetries from each region. The composite symmetry can then be used as discussed above to convert the special solution to a global one.

Treating Eqs.~\eqref{Pimugen}-\eqref{mu3soln} perturbatively in $p,\ \delta$ and $\varepsilon$, it can be seen that $n_2$ enters the $\mu\to 0$ kinetics at leading order (via $\alpha_2$). By contrast, analysis of Eq.~\eqref{genseries1} shows $n_2$ to enter the $\mu\to 1$ kinetics at second order. Thus, when the $\mu\to 0$ asymptotic region is non-vanishing, greater accuracy can be attained by transforming $\delta$ according to a $\mu\to 1$ asymptotic symmetry, but $n_2$ according to a $\mu\to 0$ asymptotic symmetry. In this way, a composite symmetry can successfully be constructed for the solution to the protein aggregation rate equations.

As above, we perform these transformations by simply replacing $c$ and $\delta$ in the special solution with functions $\tilde{c}$ and $\tilde{\delta}$ that ensure the asymptotic limit of the special solution matches that of the exact dynamics. The same $\tilde{\delta}$ is used as before since it is transformed with the same symmetry. The $\mu\to 0$ limit of the special solution Eq.~\eqref{refsoln} is:
\begin{equation}
\lim_{\mu\to 0}\mu^*(\tau,c,\delta)=e^{-c\tau}\left((1-\delta)^{-1/c}-1\right)^{-c}.
\end{equation}
The $\mu\to 0$ limit of the exact dynamics is given by Eq.~\eqref{mu3soln}. We see that to match this, $c$ must be replaced in Eq.~\eqref{refsoln} with $\tilde{c}=2k_+m_\text{tot}\Pi_\infty/\alpha_e(m_\text{tot})$, where $\Pi_\infty$ is given by Eq.~\eqref{Pimugen} with $\mu=0$. Doing so yields ultimately:
\begin{equation}
c=\frac{2k_+m_\text{tot}}{\alpha_e(m_\text{tot})}\Pi_\infty.
\end{equation}

\end{widetext}

\section{Competition between fragmentation and secondary nucleation}\label{app:competing}

Occasionally both fragmentation and secondary nucleation occur at comparable rates, whereupon it becomes useful to include both processes in a kinetic model, not just the one or the other. The rate equations are a simple generalization of Eqs.~\eqref{momeqs}:
\begin{subequations}\label{momeqsf}
	\begin{multline}\label{momeqfP}
	\frac{dP}{dt}=\frac{k_n m(t)^{n_c}}{1+\left(m(t)/K_P\right)^{n_c}}\\
	+\left(k_-+\frac{k_2 m(t)^{n_2}}{1+\left(m(t)/K_S\right)^{n_2}}\right)M(t)
	\end{multline}
	\begin{equation}\label{momeqfM}
	\frac{dM}{dt}=\frac{2k_+ m(t)}{1+m(t)/K_E}P(t)
	\end{equation}
	\begin{equation}
	m_\text{tot}=m(t)+M(t),
	\end{equation}
\end{subequations}
where $k_-M$ is the rate of fragmentation~\cite{Knowles2009}. The rate of secondary processes $\alpha_2M$ is now clearly:
\begin{equation}
\alpha_2(m)M=\left(k_-+\frac{k_2 m(t)^{n_2}}{1+\left(m(t)/K_S\right)^{n_2}}\right)M.
\end{equation}
So the key dimensionless parameters $\kappa$ and $\varepsilon$ are generalized to:
\begin{subequations}
	\begin{align}
	\kappa&=\sqrt{\frac{2k_+m_\text{tot}}{1+m_{\text{tot}}/K_E}\left(k_-+\frac{k_2m_\text{tot}^{n_2}}{1+(m_{\text{tot}}/K_S)^{n_2}}\right)}\\
	\varepsilon&=\frac{\frac{k_nm_\text{tot}^{n_c}}{1+(m_{\text{tot}}/K_P)^{n_c}}}{k_-m_\text{tot}+\frac{k_2m_\text{tot}^{n_2+1}}{1+(m_{\text{tot}}/K_S)^{n_2}}}.
	\end{align}
\end{subequations}
Their early-time equivalents are:
\begin{subequations}
	\begin{align}
	\kappa_0&=\sqrt{\frac{2k_+m(0)}{1+m(0)/K_E}\left(k_-+\frac{k_2m(0)^{n_2}}{1+(m(0)/K_S)^{n_2}}\right)}\\
	\varepsilon_0&=\frac{\frac{k_nm(0)^{n_c}}{1+(m(0)/K_P)^{n_c}}}{k_-m(0)+\frac{k_2m(0)^{n_2+1}}{1+(m(0)/K_S)^{n_2}}}.
	\end{align}
\end{subequations}
The general solution is then still Eq.~\eqref{gensolnsat} but using these modified parameters, i.e.: 
\begin{widetext}
	\begin{subequations}
	\begin{align}
	m(t)&=m(0)\left(1+\frac{p_0}{2c}\frac{\kappa_0}{\kappa}(e^{\kappa t}-e^{-\kappa t})+\frac{\kappa_0^2}{\kappa^2}\left(\frac{\varepsilon_0}{c}+\frac{\delta_0}{2c}+\frac{p_0^2}{2c^2}\left(1+\frac{c\,m(0)/K_E}{1+m(0)/K_E}\right)\right)(e^{\kappa t}+e^{-\kappa t}-2)\right)^{-c},\\
	\delta_0&=\frac{M(0)}{m(0)}, \qquad p_0=\frac{2k_+P(0)}{\kappa_0(1+m(0)/K_E)}.
	\end{align}
\end{subequations}
Owing to the presence of fragmentation the parameter $c$ should be chosen using the $\mu\to 1$ asymptotic symmetry transformation. It can again be derived using the formula Eq.~\eqref{cmu1}, i.e.:
\begin{equation}
c=\frac{3}{2n_2'+1},\quad n_2'=\frac{d\ln\!\!\left[\alpha_2(m)\alpha_e(m)^2\right]}{d\ln m}\bigg|_{m=m_\text{tot}}\!\!\!\!\!\!\!\!\!\!\!\!-2.
\end{equation}
Applying this formula gives for $n_2'$ the following, remembering that $dx/d\ln x=x$:
\begin{align}
\ln\!\!\left[\alpha_2(m)\alpha_e(m)^2\right]&=\text{const.}+2\ln m-2\ln[1+m/K_E]+\ln\!\left[k_-+\frac{k_2m^{n_2}}{1+(m/K_S)^{n_2}}\right]\\
\frac{d\ln\!\!\left[\alpha_2(m)\alpha_e(m)^2\right]}{d\ln m}&=2-\frac{2m/K_E}{1+m/K_E}+\frac{1}{k_-+\frac{k_2m^{n_2}}{1+(m/K_S)^{n_2}}}\frac{d}{d\ln m}\frac{k_2m^{n_2}}{1+(m/K_S)^{n_2}}\\
&=2-\frac{2m/K_E}{1+m/K_E}+\frac{\frac{n_2k_2m^{n_2}}{1+(m/K_S)^{n_2}}-\frac{k_2m^{n_2}}{(1+(m/K_S)^{n_2})^2}n_2(m/K_S)^{n_2}}{k_-+\frac{k_2m^{n_2}}{1+(m/K_S)^{n_2}}}\\
&=2-\frac{2m/K_E}{1+m/K_E}+\frac{\frac{k_2m^{n_2}}{1+(m/K_S)^{n_2}}}{k_-+\frac{k_2m^{n_2}}{1+(m/K_S)^{n_2}}}\cdot n_2\left(1-\frac{(m/K_S)^{n_2}}{1+(m/K_S)^{n_2}}\right).
\end{align}
Therefore:
\begin{equation}
n_2'=\frac{d\ln\!\!\left[\alpha_2(m)\alpha_e(m)^2\right]}{d\ln m}\bigg|_{m=m_\text{tot}}\!\!\!\!\!\!\!\!\!\!\!\!-2 =\frac{\frac{k_2m_\text{tot}^{n_2}}{1+(m_\text{tot}/K_S)^{n_2}}}{k_-+\frac{k_2m_\text{tot}^{n_2}}{1+(m_\text{tot}/K_S)^{n_2}}}\cdot\frac{n_2}{1+(m_\text{tot}/K_S)^{n_2}}-\frac{2m_\text{tot}/K_E}{1+m_\text{tot}/K_E}.
\end{equation}
In the limit that $k_-\to 0$ this clearly recovers Eq.~\eqref{muto1c}.

\section{Kinetics in the presence of inhibitors}\label{app:inhib}
The solution can be generalized one step further to take into account the presence of inhibitors that are present at high enough concentrations that their concentration in solution remains approximately constant. We also restrict our attention to inhibitors that bind rapidly enough to their targets that their binding can be reasonably modelled using a quasi-equilibrium approximation~\cite{Arosio2016b}.

\subsection{Monomer-binding inhibitors}
Inhibitors that bind monomeric protein directly reduce the concentration of free monomers $m_f$ that are able to take part in aggregation reaction processes. The rate equations Eqs.~\eqref{momeqsf} become:
\begin{subequations}\label{momeqsfmI}
	\begin{equation}\label{momeqfmIP}
	\frac{dP}{dt}=\frac{k_n m_f(t)^{n_c}}{1+\left(m_f(t)/K_P\right)^{n_c}}
	+\left(k_-+\frac{k_2 m_f(t)^{n_2}}{1+\left(m_f(t)/K_S\right)^{n_2}}\right)M(t)
	\end{equation}
	\begin{equation}\label{momeqfmIM}
	\frac{dM}{dt}=\frac{2k_+ m_f(t)}{1+m_f(t)/K_E}P(t)
	\end{equation}
	\begin{equation}
	m_\text{tot}=m(t)+M(t),\qquad m_f=\frac{m(t)}{1+c_d/K_I},
	\end{equation}
\end{subequations}
where $K_I$ is the dissociation constant of the inhibitor and $c_d$ its concentration. So the general solution remains the same as Eq.~\eqref{gensolnsat}, but with the following substitutions~\cite{Arosio2016b}:
\begin{subequations}
\begin{align}
\left\lbrace k_+,k_n,k_2\right\rbrace &\to \left\lbrace\frac{k_+}{1+c_d/K_I},\frac{k_n}{(1+c_d/K_I)^{n_c}},\frac{k_2}{(1+c_d/K_I)^{n_2}}\right\rbrace \\
\left\lbrace K_E,K_P,K_S\right\rbrace &\to \left\lbrace K_E(1+c_d/K_I),K_P(1+c_d/K_I),K_S(1+c_d/K_I)\right\rbrace.
\end{align}
\end{subequations}

\subsection{Oligomer-binding inhibitors}
If an on-pathway oligomer is bound and this oligomer is an intermediate of both primary and secondary nucleation, then its conversion rate is effectively reduced, but not its formation rate. So, this effectively reduces the rate constants of primary and secondary nucleation. If the oligomer is intermediate of only one of these nucleation steps, then only one of the nucleation rate constants is modified. The general solution therefore applies, with the following simple modifications:
\begin{subequations}
\begin{align}
\text{If inhibitor binds intermediate of primary nucleation: } k_n &\to \frac{k_n}{1+c_d/K_I} \\
\text{If inhibitor binds intermediate of secondary nucleation: } k_2 &\to \frac{k_2}{1+c_d/K_I}.
\end{align}
\end{subequations}

\subsection{Competitive surface-binding inhibitors}
When the inhibitor binds the fibril ends, the secondary nucleation sites on the fibril surface, or the interface where primary nucleation occurs, then it competes with monomer for the binding site. For secondary nucleation this phenomenon was explored in~\cite{Dear2023B}. A modified rate law for secondary nucleation was derived describing this competition:
\begin{equation}
\alpha_2(m)M=\frac{k_2m^{n_2}M}{1+(m/K_S)^{n_2}+c_d/K_{I,S}},
\end{equation}
where $K_{I,S}$ is the dissociation constant for inhibitor from secondary nucleation sites. It is trivial to modify this to a combined rate law for secondary nucleation and fragmentation, and also to modify the rate laws for primary nucleation and elongation to account for competitive inhibition in the same way. The overall rate equations become:
\begin{subequations}\label{momeqsfcI}
	\begin{equation}\label{momeqfcIP}
	\frac{dP}{dt}=\frac{k_n m(t)^{n_c}}{1+\left(m(t)/K_P\right)^{n_c}+c_d/K_{I,P}}
	+\left(k_-+\frac{k_2 m(t)^{n_2}}{1+\left(m(t)/K_S\right)^{n_2}+c_d/K_{I,S}}\right)M(t)
	\end{equation}
	\begin{equation}\label{momeqfcIM}
	\frac{dM}{dt}=\frac{2k_+ m(t)}{1+m(t)/K_E+c_d/K_{I,E}}P(t)
	\end{equation}
	\begin{equation}
	m_\text{tot}=m(t)+M(t),
	\end{equation}
\end{subequations}
where $K_{I,E}$ and $K_{I,P}$ are the dissociation constants for inhibitor from fibril ends and from secondary nucleation sites respectively. 

Once more, the general solution is still given by Eq.~\eqref{gensolnsat} but with $1+m(0)/K_E$ replaced by $1+m(0)/K_E+c_d/K_{I,E}$ and using similarly modified parameters and Eq.~\eqref{Q}, i.e.: 
\begin{subequations}
	\begin{align}
	m(t)&=m(0)\left(1+\frac{p_0}{2c}\frac{\kappa_0}{\kappa}(e^{\kappa t}-e^{-\kappa t})+\frac{\kappa_0^2}{\kappa^2}\left(\frac{\varepsilon_0}{c}+\frac{\delta_0}{2c}+\frac{p_0^2}{2c^2}\left(1+\frac{c\,m(0)/K_E}{1+m(0)/K_E+c_d/K_{I,E}}\right)\right)(e^{\kappa t}+e^{-\kappa t}-2)\right)^{-c},\\
	\delta_0&=\frac{M(0)}{m(0)}, \qquad p_0=\frac{2k_+P(0)}{\kappa_0(1+m(0)/K_E+c_d/K_{I,E})},
	\end{align}
\end{subequations}
where the key dimensionless parameters $\kappa,\ \varepsilon$ and $\kappa_0,\ \varepsilon_0$ are now:
\begin{subequations}
	\begin{align}
	\kappa&=\sqrt{\frac{2k_+m_\text{tot}}{1+m_{\text{tot}}/K_E+c_d/K_{I,E}}\left(k_-+\frac{k_2m_\text{tot}^{n_2}}{1+(m_{\text{tot}}/K_S)^{n_2}+c_d/K_{I,S}}\right)}\\
	\varepsilon&=\frac{\frac{k_nm_\text{tot}^{n_c}}{1+(m_{\text{tot}}/K_P)^{n_c}+c_d/K_{I,P}}}{k_-m_\text{tot}+\frac{k_2m_\text{tot}^{n_2+1}}{1+(m_{\text{tot}}/K_S)^{n_2}+c_d/K_{I,S}}}\\
	\kappa_0&=\sqrt{\frac{2k_+m(0)}{1+m(0)/K_E+c_d/K_{I,E}}\left(k_-+\frac{k_2m(0)^{n_2}}{1+(m(0)/K_S)^{n_2}+c_d/K_{I,S}}\right)}\\
	\varepsilon_0&=\frac{\frac{k_nm(0)^{n_c}}{1+(m(0)/K_P)^{n_c}+c_d/K_{I,P}}}{k_-m(0)+\frac{k_2m(0)^{n_2+1}}{1+(m(0)/K_S)^{n_2}+c_d/K_{I,S}}}.
	\end{align}
\end{subequations}
We can use again the $\mu\to 1$ asymptotic symmetry transformation formula Eq.~\eqref{cmu1} for $c$, i.e.:
\begin{equation}
c=\frac{3}{2n_2'+1},\quad n_2'=\frac{d\ln\!\!\left[\alpha_2(m)\alpha_e(m)^2\right]}{d\ln m}\bigg|_{m=m_\text{tot}}\!\!\!\!\!\!\!\!\!\!\!\!-2.
\end{equation}
$n_2'$ is derived using this formula in the same way as before, yielding:
\begin{align}
\ln\!\!\left[\alpha_2(m)\alpha_e(m)^2\right]&=\text{const.}+2\ln m-2\ln[1+m/K_E+c_d/K_{I,E}]+\ln\!\left[k_-+\frac{k_2m^{n_2}}{1+(m/K_S)^{n_2}+c_d/K_{I,S}}\right]\\
\frac{d\ln\!\!\left[\alpha_2(m)\alpha_e(m)^2\right]}{d\ln m}&=2-\frac{2m/K_E}{1+m/K_E+c_d/K_{I,E}}+\frac{1}{k_-+\frac{k_2m^{n_2}}{1+(m/K_S)^{n_2}+c_d/K_{I,S}}}\frac{d}{d\ln m}\frac{k_2m^{n_2}}{1+(m/K_S)^{n_2}+c_d/K_{I,S}}\\
&=2-\frac{2m/K_E}{1+m/K_E+c_d/K_{I,E}}+\frac{\frac{n_2k_2m^{n_2}}{1+(m/K_S)^{n_2}+c_d/K_{I,S}}-\frac{k_2m^{n_2}}{(1+(m/K_S)^{n_2}+c_d/K_{I,S})^2}n_2(m/K_S)^{n_2}}{k_-+\frac{k_2m^{n_2}}{1+(m/K_S)^{n_2}+c_d/K_{I,S}}}\\
&=2-\frac{2m/K_E}{1+m/K_E+c_d/K_{I,E}}+\frac{\frac{k_2m^{n_2}}{1+(m/K_S)^{n_2}+c_d/K_{I,S}}}{k_-+\frac{k_2m^{n_2}}{1+(m/K_S)^{n_2}+c_d/K_{I,S}}}\cdot n_2\left(1-\frac{(m/K_S)^{n_2}}{1+(m/K_S)^{n_2}+c_d/K_{I,S}}\right).
\end{align}
Therefore:
\begin{equation}
n_2'=\frac{\frac{k_2m_\text{tot}^{n_2}}{1+(m_\text{tot}/K_S)^{n_2}+c_d/K_{I,S}}}{k_-+\frac{k_2m_\text{tot}^{n_2}}{1+(m_\text{tot}/K_S)^{n_2}+c_d/K_{I,S}}}\cdot\frac{n_2(1+c_d/K_{I,S})}{1+(m_\text{tot}/K_S)^{n_2}+c_d/K_{I,S}}-\frac{2m_\text{tot}/K_E}{1+m_\text{tot}/K_E+c_d/K_{I,E}}.
\end{equation}

\end{widetext}

\bibliography{bibliography.bib}
\end{document}